# Interplay of valley, layer and band topology towards interacting quantum phases in moiré bilayer graphene

Yungi Jeong[1,2], Hangyeol Park[1,2], Taeho Kim[2], K. Watanabe[3], T. Taniguchi[4], Jeil Jung[5,6], Joonho Jang[1,2]*

[1] Center for Correlated Electron Systems, Institute for Basic Science, Seoul 08826, Korea

[2] Department of Physics and Astronomy, and Institute of Applied Physics, Seoul National University, Seoul 08826, Korea

[3] Research Center for Electronic and Optical Materials, National Institute for Materials Science, 1-1 Namiki, Tsukuba 305-0044, Japan

[4] Research Center for Materials Nanoarchitectonics, National Institute for Materials Science, 1-1 Namiki, Tsukuba 305-0044, Japan

[5] Department of Physics, University of Seoul, Seoul, Korea

[6] Department of Smart Cities, University of Seoul, Seoul, Korea

*Corresponding author's e-mail: joonho.jang@snu.ac.kr

**Abstract: In Bernal-stacked bilayer graphene (BBG), the Landau levels give rise to an intimate connection between valley and layer degrees of freedom. Adding a moiré superlattice potential enriches the BBG physics with the formation of topological minibands — potentially leading to tunable exotic quantum transport. Here, we present magnetotransport measurements of a high-quality bilayer graphene–hexagonal boron nitride (hBN) heterostructure. The zero-degree alignment generates a strong moiré superlattice potential for the electrons in BBG and the resulting Landau fan diagram of longitudinal and Hall resistance displays a Hofstadter butterfly pattern with a high level of detail. We demonstrate that the intricate relationship between valley and layer degrees of freedom controls the topology of moiré-induced bands, significantly influencing the energetics of interacting quantum phases in the BBG superlattice. We further observe signatures of field-induced correlated insulators, helical edge states and clear quantizations of interaction-driven topological quantum phases, such as symmetry broken Chern insulators.**

## Introduction

Bernal bilayer graphene (BBG) is the simplest member of the rhombohedral stacked multilayer graphene family. By breaking the inversion symmetry of BBG with a perpendicular electric displacement field $D$, BBG shows a tunable gap opening and van Hove singularities which can lead to a cascade of symmetry broken states.[1–4] Also, electrons in BBG have a layer degree of freedom correlated with the valley degree of freedom in a specific way, which makes it possible to investigate the proximity effect of substrates surrounding BBG. When BBG is aligned with hexagonal boron nitride (hBN), the slight lattice mismatch (~1.8%) of the layers creates a periodic moiré superlattice potential. Especially in this system, the zero-degree rotation between BBG and hBN is known to be the most stable with the global configurational energy minimum; thus it is expected to have very low superlattice disorder. This superlattice potential is expected to modify the band structure near the K and K' points and induces secondary Dirac points and flat bands with non-trivial topological properties.[5] Remarkably, in the presence of a perpendicular magnetic field, the so-called Hofstadter's butterfly appears due to the interplay between superlattice potential and magnetic field.[6-9] When the magnetic flux per superlattice unit cell becomes a rational multiple of the flux quantum, a commensurability between the cyclotron orbits and the lattice periodicity leads to regain a lattice translational symmetry to form extended electronic states known as Brown–Zak (BZ) quasiparticles.[10-13] These BZ quasiparticles, satisfying the new Bloch equation, move through the lattice as if the magnetic field is zero. Away from these commensurable magnetic field values, BZ quasiparticle feels an effective magnetic field $B_{eff} = B - B_{p/q}$ (penetration of p flux quanta per q superlattice unit cells.) and form mini Landau fans radiating from $B = B_{p/q}$ lines. Because BZ quasiparticles can be defined for each rational $p/q$, a fractal-like self-repeating structure - called the Hofstadter butterfly[14-19] - would emerge.

In addition to single-particle effects, electron–electron interactions can play a significant role in these systems due to the relatively flat moiré bands, resulting in large density of states at certain energies. These interactions lead to quantum correlated phases, and their interplay with the band topology adds another layer of controllability and richness to the physics of BBG/hBN systems. By tuning the $D$ field, which affects the interlayer potential, one expects to control the topological properties and electrical energy bandwidth of the system, offering a pathway to explore a range of interacting topological quantum states and their phase transitions, and potentially realize devices where the topological property is dynamically controlled. Understanding of the interplay between the valley and layer physics of BBG and the superlattice potential from hBN thus is critical. However, despite the desirable combination of the stable configuration and electric- and magnetic-field tunability, the electron transport properties in a BBG/hBN moiré superlattice are relatively less explored compared to the other moiré systems, mostly due to the technical difficulty in making high-quality samples with both top- and bottom-gate electrodes and robust ohmic transport contacts necessary in a strong magnetic field and superlattice potential. In this article, we report a magneto-transport study of a high mobility zero-degree aligned BBG/hBN heterostructure with dual graphite gates and four high-transparent contacts that allow successful longitudinal and Hall measurements. The high quality of the sample and the zero degree alignment allow us to observe a large variety of Chern insulator states and interaction driven states even under relatively low magnetic fields, revealing an intricate interplay of various quantum degrees of freedom.

## Results

### Device characterization

BBG/hBN aligned sample was fabricated by a conventional dry-transfer method using PDMS and PC film[20] (**Fig. 1a**). **Fig. 1b** shows the stacking structure of the dual gated BBG/hBN aligned device. Since conventional metal contacts can open gaps in the BBG[21,22], an additional graphite layer was used to ensure that contact resistance remains sufficiently low even when strong magnetic and displacement fields are applied in cryogenic temperatures. We performed four-probe electrical transport measurements of the sample. **Fig. 1c** shows the resistance as a function of carrier density at zero magnetic and displacement field. Due to the superlattice potential imposed by the lattice mismatch between hBN and BBG, induced energy gap or DOS minimum is expected at the energies below and above the charge neutral point (CNP)[7,8]. Actually, a recent experiment supports DOS minimum rather than full gap[23] (see also Supplementary Note 9). We indeed observe two satellite peaks on either side of CNP at positive and negative densities, $n = \pm 2.397 \times 10^{12}$ cm$^{-2}$ ( $= \pm 4n_0$), where the superlattice-induced isolated energy bands are either completely full or empty. From the value, we estimate that the lattice constant of the superlattice is 13.88 nm and determine that the twist angle between the BBG and the top hBN is 0 degree within our experimental accuracy. **Fig. 1d** shows the Hall mobility as a function of carrier density, and the mobility near the CNP is about 200,000 cm$^2$ V s$^{-1}$ comparable to high mobility GaAs[24–26] or suspended graphene[27,28], indicating that the sample is of very high quality. Unlike the free-standing BBG, we notice that the mobility dips near the full-filling the isolated superlattice bands, probably due to the small number of effective charge carriers near the full filling of moiré band.

**Magnetotransport measurement**

**Fig. 2** shows longitudinal and Hall resistance measurements in the presence of a perpendicular magnetic field (see also **Supplementary Figure 1, 2** for various $D$ field values). In **Fig. 2a**, the dark regions of low longitudinal resistance are identified as incompressible states with edge states. The effects of the superlattice in the spectrum are readily noticeable; at the intersections between the horizontal lines at $\phi/\phi_0 = p/q$ and lines of the conventional integer quantum Hall states, mini fans appear to form the fractal-like Hofstadter's butterfly. In the Hall resistance measurements in **Fig. 2c**, we directly identify the effective magnetic field $B_{eff} = B - B_{p/q}$ felt by BZ quasiparticles with sign changes of Hall resistance at $\phi/\phi_0 = p/q$, forming a horizontal pattern of white strips. We also find the $D$ field induced insulating gap at CNP abruptly closes near $B = 9$ T due to the superlattice effects (green arrow in **Fig. 2a**; see also **Supplementary Figure 3**). In the low magnetic field region near CNP, the Landau fan looks qualitatively similar to the case of an intrinsic BBG with the spin–valley subbands of Landau levels resolved even below $B = 1$ T but, in higher magnetic fields, the splittings become less distinct and even disappear, due to the dominant effect of the moiré potential leading to significant overlaps between LL subbands, while ushering in the appearance of moiré-induced incompressible states.

Each incompressible state in the Landau fan spectrum follows the Diophantine equation $n/n_0 = t(\phi/\phi_0) + s$, where the slope t gives the total Chern number of occupied bands proportional to the Hall conductance, and the n-intercept s gives the number of charges trapped in the superlattice unit cell. Unlike the conventional quantum Hall states ($s = 0$), nonzero s states have charge density and Hall conductance decoupled by a strong superlattice potential.[19,29] To investigate these states thoroughly, we plotted a Wannier diagram in **Fig. 2b**, where the light gray lines represent all possible trajectories allowed by the Diophantine equation. Then, we overlaid additional colored lines that correspond to the observed insulators in **Fig. 2a**. First, the dark gray lines correspond to the conventional integer quantum Hall effect (IQHE) (integers $t$ and $s = 0$), and the black lines identify the Chern insulators (CI) (integer $t$, integer $s \neq 0$); all these states can appear in a non-interacting single-particle Hofstadter spectrum. On the other hand, due to particle interaction, more incompressible features appear in the data especially when measured at $T = 30$ mK. For example, the green lines indicate the fractional quantum Hall effect (FQHE) (fractional $t$, $s = 0$), where the well-known FQHE at $\nu = m - 1/3$ ($m = \pm 1, \pm 3$) [30] are visible as low as $B = 5\,\text{T}$, reflecting the high quality of the electron system. And several red lines denote symmetry broken Chern insulator (SBCI) (integer $t$, fractional $s$) whose trajectories are located at fractional filling of moiré Chern bands. We emphasize that most of the Chern numbers of the incompressible states are independently confirmed by the Hall quantization measurements.

### $D$ field tunable valley selective moiré effect

The electronic system displays dramatic changes in spectra when tuned with $D$ field, due to BBG's layer degrees of freedom and its peculiar property associated with the valley degrees of freedom. In particular, striking bright fork-like features appear and strongly depend on $D$ field (red arrows in **Fig. 2a, Supplementary Figure 1**). These fork-like features, when plotted as a function of $D$ and $n$ as in **Fig. 3a–c**, appear as thick bright lines. Interestingly, for higher Landau levels (LLs) with $N \geq 2$ (**Fig. 3a–b**), the bright lines move in the opposite direction to the case for the zero-energy Landau levels (ZLLs) (**Fig. 3c, Supplementary Figure 4**). This phenomenon is depicted in the schematic of **Fig. 3d**. To explain this observation, we consider the energies of the LLs of BBG given by[32],

$$\begin{cases} \epsilon_0 = \frac{1}{2}\xi U + E_s\sigma \\ \epsilon_1 = \left(\frac{1}{2} - \frac{\hbar\omega_c}{\gamma_1}\right)\xi U + E_s\sigma + \Delta_{10} \\ \epsilon_N^{\pm} = \pm\hbar\omega_c\sqrt{N(N-1)} - \frac{\hbar\omega_c}{2\gamma_1}\xi U + E_s\sigma \ (N \geq 2) \end{cases} \quad (1)$$

Here, $N$ is the orbital index of Landau levels, $\omega_c$ is the cyclotron frequency of carriers in BBG, $\xi$ is the valley index (+1 for K valley, -1 for K' valley), $\gamma_1$ is the interlayer hopping parameter[33,34], $U$ is the interlayer potential given as $U = D\ d_0$ ($d_0$: interlayer distance of BBG), $E_s$ is the spin-dependent energy splitting, that includes the Zeeman energy, to lift spin degeneracy of the bands, and $\Delta_{10} \propto \gamma_4 B$ ($\gamma_4$ is the skew tunneling term from non-dimer carbon site to dimer carbon site[33,34]) is an additional energy difference between $N = 0$ and 1.

Because $\hbar\omega_c/\gamma_1$ is smaller than $1/2$ for the range of magnetic fields used in our experiment, $\epsilon_0$, $\epsilon_1$ for ZLLs ($N = 0,1$) and $\epsilon_N$ for higher LLs ($N \geq 2$) have the opposite coefficients of energy shifts upon the change of $U$ (and thus $D$) for a specific valley; i.e. for a fixed valley, the layer polarization $p = -\partial\epsilon/\partial D$ of a higher LL has the opposite sign to the one of a ZLL. Thus, the energy of the K (K') valley should increase (decrease) with increasing $D$ field in ZLLs, while it decreases (increase) in higher LLs. The schematic in **Fig. 3e** based on this valley analysis fully explains the behavior of the bright features in **Fig. 3a–c**. It thus strongly suggests that the two out of the four symmetry-broken subbands of a LL selectively experience the stronger superlattice effect and that all the bright bands are actually of a specific valley (denoted as K' valley for both ZLLs and LLs with $N \geq 2$) experiencing more scattering that results in a higher resistance. We point out that this finding is somewhat counterintuitive and contrary to the previous beliefs[7,8] in that the degrees of freedom that determine the effect of the moiré potential are not the layers but actually the valleys (**Fig. 3d–e**). We further performed numerical simulation (**Supplementary Figure 15**) that supports the phenomenology of significant subband broadenings for one of the valleys. Interestingly, the topology of the superlattice-induced isolated bands play an important role here; the combined Chern number of the induced isolated bands has the exact opposite value for each valley as shown in **Fig. 3f** (see also **Supplementary Figure 16**), and we find that the observed valley-selective moiré effect originates from the disparate magnetic field responses of the isolated bands with opposite Chern numbers because the band edges merge at lower magnetic fields for the isolated band (of K' valley) with negative Chern number, inducing more severe modification of LL spectra[5,35,36]. More discussions on the potential microscopic mechanism of this phenomena are in Supplementary Note 6.

The gate tunability further extends to the topological properties of the moiré bands. By applying $D$ field, we control the valley-selective effect of moiré potential, indirectly via the layer polarization, whose values are peculiarly intertwined with the valleys. In particular, various Chern insulator states are controlled due to the interplay of valleys and layers upon varying the vertical $D$ field. In **Fig. 3g–h**, we plotted data showing the $D$ field tunability of the $N = 2$ LL. Inside each LL subbands, multiple Chern gaps appear and disappear strongly depending on $D$ field values. We find that the Chern insulators transit under a certain rule: while IQHE gap at $\nu = 5, 6, 7$ closes with LL subbands switching their positions in filling sequence, the Chern insulators whose Chern numbers $t$ differ by 1 (or -1) but with the same $s$ values appear in the adjacent LL subbands in a cascading fashion (**Supplementary Figure 5**) This suggests the Chern insulators are valley- and spin-polarized, and adds to the idea that tuning valley or spin degrees of freedom of the Chern insulators is a key to control topology in this BBG system.

**Correlated insulating states**

Another outstanding feature in the spectra is the existence of the insulating states located at $n/n_0 = -1, -2$ surprisingly persistent throughout the values of B, as shown in **Fig. 4**. At $B = 0$, there is an insulating phase at $n/n_0 = 0$, but upon increasing the magnetic field, new insulators develop at $n/n_0 = $ -1, -2 (blue arrows in **Fig. 4a**). We evaluated the bulk insulating gap of (0,-2) state by fitting the temperature dependence to the Arrhenius formula (**Fig. 4b**), and found the gap size is particularly larger than other nearby superlattice-induced Chern insulating states. This is not explainable by our simulation based only on the single-particle picture (**Supplementary Figure 14**), and thus strongly suggests that electron correlation strongly enhances the energy gaps of the states. Such a correlated insulator can emerge when particle interaction leads to spontaneous spin or valley order and saves energy by filling the superlattice with one particle per unit cell[37]. We attribute this strong correlation to the narrow bandwidth of the isolated valence band between the CNP (at $n = 0$) and superlattice induced insulator at $n/n_0 = -4$. According to the simulation (**Supplementary Figure 13**), the bandwidth is to be smaller than 50 meV and decreases upon increasing the strength of D. Certainly, most of the fractional states and correlated insulating states we have observed exist between $n/n_0 = 0$ and $-4$, suggestive of the narrow moiré-induced isolated band playing an important role in enhancing the interaction effects. In addition, we changed the configuration of the contact as in **Supplementary Figure 8b** to measure the non-local resistance ($R_{NL}$) of the sample and found that $R_{NL}$ is significantly large at (0,-1). We attribute the high non-local resistance of the state to a helical edge channel of counter-propagating valleys (see Supplementary Note 9). However, we cannot rule out the complicated involvement of spins in the edge channel, such as one in a spin–valley magnetic state. A measurement with a large in-plane magnetic field to control Zeeman energy may help to resolve the question while it is out of the scope of the current work.

## Fractional incompressible states

The effect of particle interaction due to the narrow bandwidth (**Fig. 4**) and tunable band topology (**Fig. 3**) further leads to exotic incompressible fractional states. Remarkably, in **Fig. 5a–c** and **Supplementary Figure 7**, we clearly identify multiple phases whose values of s are fractional even below B = 7T, which is 2–3 times lower than previous reports[29,38,39] and another manifestation of the high quality of this sample. These observed interaction driven states have gap trajectories described by $(t, s) = (t_L, s_L) + \nu_C(\Delta t = t_R - t_L, \Delta s = s_R - s_L)$ (where $(t_{L,R}, s_{L,R})$ is the neighboring left and right integer states respectively) with a fractional $\nu_C$. We observed states with $\nu_C$=1/2 in bands with Chern number $\Delta t$=-2 ((-5,1/2), (-3,1/2)) and states with $\nu_C$=1/3 and 2/3 in a band with $\Delta t$=3 ((-5,2/3), (-4,1/3)). These (integer $t$, fractional $s$) states are interpreted as symmetry broken Chern insulators (SBCIs) arising from the spontaneous doubling/tripling of the superlattice unit cell because they do not follow the theoretically expected filling factor $\nu_C$ of fractional Chern insulators (FCIs) in a Chern band with $|\Delta t| > 1$.[40,41] However, we still cannot exclude the possibility of partial filling of the moiré Chern band with exotic fractionalized excitations. The techniques that can detect charge quantization, such as shot noise measurements[42], would be helpful to resolve this issue. In **Fig. 5e–h**, the energy gaps of the SBCI states were extracted by fitting to the Arrhenius formula. We can divide these four states into two relatively large (**Fig. 5e, h**) and two relatively small (**Fig. 5f, g**) gap pairs. These two pairs have different Chern numbers of the fractal bands supporting each state, which suggests that the band topology is closely related to the nature of the fractional states (see also **Supplementary Figure 7f–g**). Also, we observed that unlike the conventional FQH states in a misaligned BBG appearing in a relatively wide range of $D$ field[30,43,44], fractional states in our sample exist only in a narrow range of $D$ field and seem to critically depend on the fractal band's bandwidth and topology, defined by the band Chern number.

## Discussion

In conclusion, we have shown the $D$ field dependent interacting Hofstadter spectrum in BBG/hBN moiré system and discovered that, upon application of $D$ field, the system's valleys respond according to their LL-dependent layer polarization and generate distinct spectral features that strongly depend on the valley-dependent band Chern number. Also, we experimentally identified interaction-driven states in the fractal bands based on their definitive fractional Hall quantization. Our work demonstrates that owing to the in-situ tunability with gate electrodes and the high-quality assisted by its mechanical stability, BBG/hBN moiré system provides a highly-tunable platform for studying the interplay of band topology and electron correlation, and opens up exciting opportunities to explore spin–valley isospin polarizations of interacting states in Hofstadter spectrum. Further theoretical investigations on energetics of SBCIs and FCIs and its quantitative relationship to the Chern number of fractal band would be highly desired.

**Methods**

**Device fabrication**

Encapsulated hBN (aligned)/BBG/hBN device was fabricated using the van der Waals dry transfer technique. The entire stack was picked up by PDMS/PC stamp in the following order: hBN (for graphite pickup), top graphite gate, top hBN (34 nm), contact graphite , BBG, bottom hBN (61 nm), bottom graphite gate. Graphite flakes were used only after ensuring that they had at least 7–8 layers by optical contrast. The stack was dropped down on the pre-defined align marker pattern and annealed in a vacuum furnace 500 degrees for 2 hours to accumulate small bubbles in the stack. After electron beam lithography, aluminum was deposited and used as an etch mask, and since the contact graphite is only in contact with one side of the BBG, the sample is etched in the shape of a horseshoe rather than a Hall bar. Gate and contact graphite were edge contacted with Ti/Au metallic leads. Finally, just before measurement, the sample was annealed in a vacuum furnace at 400 degrees for 1 hour.

**Measurement**

Measurements were taken in a cryogen-free dilution refrigerator with a base temperature of 20 mK. Due to the slight temperature increase caused by operating the superconducting magnet, most of the actual measurements were performed at 30 mK. An RC/RF electric filter and a sapphire stripline heat sink[45] were used to lower the electron temperature. Electrical measurements were performed using standard lock-in amplifier techniques. To measure longitudinal resistance, an AC voltage bias of $1$ mV$_{\text{RMS}}$, 13.33 Hz was connected to the top contact of the device in series with a 1 M$\Omega$ resistor to simulate an AC current bias of $1$ nA$_{\text{RMS}}$. The lowest contact was then connected to the current input of the SR865A lock-in amplifier to measure the ac current (**Fig.1a**). Voltage was measured between the second and third contacts in the middle of the device using a SR560 voltage preamplifier with a gain of 1000. Then the longitudinal resistance was defined as $R_{xx} = V/I$. Due to the large parameter space (carrier density, displacement field, magnetic field, temperature), most of the measurements were performed by one-shot measurement or by taking a small number of measurements and averaging them to reduce the measurement time. However, if the parameter sweep speed is not set appropriately, it can cause measurement errors, so it is important to set the appropriate saturation time according to the measurement frequency and the time constant of the amplifier.

**Numerical simulation**

We performed a numerical simulation of the electronic energy spectra in a BBG/hBN with zero-degree alignment by the direct diagonalization of a Hamiltonian matrix following the continuum model proposed by Bistritzer and Macdonald[46]. Due to the natural lattice mismatch of 1.08%, electrons in the BBG experience a superlattice potential imposed at the interface to the hBN even when the twisted angle is zero, which is known to be the energetically stable configuration after considering the lattice relaxation.

At a finite perpendicular magnetic field, the following Hamiltonian becomes numerically solvable when expressed in the bases of Landau gauge eigenfunctions (usually truncated at LLs above $N$~200 for the manageable computation time, but still with a good approximation).

$$\hat{H} = \begin{pmatrix} (-U-\Delta_{sub})/2 & -\pi & v_4\pi & v_3\pi^\dagger & 0 & 0 \\ -\pi^\dagger & (-U+\Delta_{sub})/2 & \gamma_1 & v_4\pi & 0 & 0 \\ v_4\pi^\dagger & \gamma_1 & (U-\Delta_{sub})/2 & -\pi & & T \\ v_3\pi & v_4\pi^\dagger & -\pi^\dagger & (U+\Delta_{sub})/2 & & \\ 0 & 0 & & T^\dagger & U_B & 0 \\ 0 & 0 & & & 0 & U_N \end{pmatrix} \quad (2)$$

$$\text{(where } \pi \equiv \xi p_x - i p_y, \ p_i \equiv -i\hbar\nabla_i - eA_i\text{)}$$

, where it is written in the bases of $|\psi> = (|A1>, |B1>, |A2>, |B2>, \underline{|B>}, \underline{|N>})$. Here, $|A1>$ and $|B1>$ are the bases of the top graphene sublattices, $|A2>$ and $|B2>$ are of the bottom graphene, and $\underline{|B>}$ and $\underline{|N>}$ are the atomic Boron and Nitrogen subbases of hBN, respectively. Then, we have used the following parameters in the simulation: $v = 9.1 \times 10^7$ cm/s , $\gamma_1 = 400$ meV , $v_3 = 9.0 \times 10^6$ cm/s , $v_4 = 4.5 \times 10^6$ cm/s , $U_B = -1400$ meV and $U_N = +3300$ meV[7,47,48]. We set $\Delta_{sub} = 0$ meV. The matrix $T$ represents the interlayer moiré hopping between the bottom graphene and hBN and is written as,

$$T = \left[ \begin{pmatrix} u & u' \\ u' & u \end{pmatrix} + e^{i\xi \mathbf{G'_1}\cdot\mathbf{r}} \begin{pmatrix} u & u'\omega^{-\xi} \\ u'\omega^\xi & u \end{pmatrix} + e^{i\xi(\mathbf{G'_1}+\mathbf{G'_2})\cdot\mathbf{r}} \begin{pmatrix} u & u'\omega^\xi \\ u'\omega^{-\xi} & u \end{pmatrix} \right] \quad (3)$$

, where we used two different parameters $u' = 130$ meV, $u = 0.8 \times 130$ meV to account for the lattice relaxation effect[49,50], and $\omega = e^{2\pi i/3}$. $\mathbf{G}$'s are the reciprocal lattice vectors of the lattice-mismatch-induced moiré potential. Note the four blocks boxed with dotted-lines hybridize two monolayer graphene blocks and a hBN block, so that the Hamiltonian represents the whole BBG/hBN heterostructure. By setting the blocks labeled as $T$ and $T^\dagger$ to 2-by-2 null matrices, one recovers the intrinsic BBG spectra.

**Data availability**

The source data used in this study are available in the figshare database under accession code https://doi.org/10.6084/m9.figshare.24119286. Other data that support the findings of this study are available from the corresponding author upon request.

**Code availability**

The codes related to the findings of this study are available from the corresponding authors upon request.

**Acknowledgements**

We gratefully acknowledge helpful discussions with B. J. Yang. The work at SNU was supported by the National Research Foundation of Korea grants funded by the Ministry of Science and ICT (Grant Nos. 2019R1C1C1006520, 2020R1A5A1016518, RS-2023-00258359), the Institute for Basic Science of Korea (Grant No. IBS-R009-D1), SNU Core Center for Physical Property Measurements at Extreme Physical Conditions (Grant No. 2021R1A6C101B418), Creative-Pioneering Researcher Program through Seoul National University and Samsung DS Basic Research Program (Project No. 0409-20230298). J. J. acknowledges support from Samsung Science and Technology Foundation Grant No. SSTF-BA1802-06. K. W. and T. T. acknowledge support from the JSPS KAKENHI (Grant Numbers 21H05233 and 23H02052) and World Premier International Research Center Initiative (WPI), MEXT, Japan.

**Author Contributions**

Y.J. and J.Jang conceived the project, Y.J. fabricated the device and performed measurements with the help from H.P. and T.K. Y.J. and J.Jang analyzed data and performed numerical calculations with the help from J.Jung. K.W. and T.T. grew the single crystal hBN. Y.J., J.Jung and J.Jang wrote the manuscript with inputs from all authors. J.Jang supervised the overall project.

**Competing interests**
The authors declare no competing interests.

**Figure Captions**

**Fig. 1 | Device geometry and zero magnetic field transport measurements**. **a,** Optical microscope image of the aligned BBG/hBN heterostructure device. Scale bar, 15 μm. The crystalline axis of bottom hBN and BBG was aligned at almost 0 degrees. Inset: Schematic of the moiré pattern seen in the 0 degree aligned BBG/hBN device. **b,** Schematic diagram of the BBG/hBN aligned device. gr denotes the graphite flake. Top and bottom gates make it possible to tune the carrier density and $D$ field simultaneously. And the graphite contact layer was additionally inserted to make a good ohmic contact to the BBG. **c,** Longitudinal resistance versus carrier density at zero magnetic and displacement field. Satellite peaks were observed at $n = \pm 2.397 \times 10^{12}/$ cm$^2$ on either side of the CNP peak. **d,** Hall mobility versus carrier density at zero magnetic and displacement field. Mobility decreases as the carrier density approaches the superlattice full fillings from CNP.

**Fig. 2 | Magnetotransport in a BBG/hBN moiré superlattice**. **a,** Longitudinal resistance, as a function of normalized carrier density and magnetic flux, measured at $T = 30$ mK with $D = 91$ mV nm$^{-1}$. Here, n is the carrier density, $n_0$ is the density corresponding to full filling of superlattice, $\phi$ is the magnetic flux per superlattice unit cell and $\phi_0 = h/e$ is the magnetic flux quantum. The red arrows point to some Landau levels that are strongly influenced by the moiré potential and exhibit high resistance. The green arrow indicates the point where the bandgap at CNP closes under the effect of moiré potential. The color scale is truncated at 10 k$\Omega$. **b,** A Wannier diagram to denote observed incompressible states identified in **a**. We assume that both spin and valley degeneracies are lifted, and only show up to $|t| \leq 8$. Light gray lines indicate incompressible states allowed by Diophantine equation assuming spin and valley degeneracy are lifted. Four classes of trajectories are distinguished by color: Integer quantum Hall insulators (gray; integer $t$, $s = 0$), Fractional Quantum Hall insulators (green; fractional $t$, $s = 0$), moiré potential induced Chern insulators (black; integer $t$, integer $s$, $s \neq 0$), Symmetry broken Chern insulators (red; integer $t$, fractional $s$). The blue and orange horizontal lines represent the first and second order Brown–Zak oscillations ($\phi/\phi_0 = 1/q, 2/q$), respectively. **c,** Hall resistance data in the hole-doped region at the same temperature and $D$ field.

**Fig. 3 | Degeneracy lifted Landau levels with moiré superlattice potential. a–c,** Longitudinal resistance $R_{xx}$ as a function of carrier density and $D$ field at $B =$ **a,** 3.8 T, **b,** 5.5 T, and **c,** 8.23 T ($\phi/\phi_0 = 1/3$) at $T = 30$ mK. N denotes the orbital index of the Landau level (LL) of BBG, and white-colored numbers indicate the Landau filling factors ν of integer quantum Hall states. Integer quantum Hall states appear dark due to its dissipationless edge state, while metallic states that partially fill the LL have relatively high resistance and appear bright. **d,** Schematic of filling sequences in **a–c.** l (lower) and u (upper) denote the layer polarization of each state. **e,** Schematic of LL energy spectrum. Each state is denoted by $|N\xi\sigma\rangle$ (orbital, valley, and spin). The tendency of the energy with respect to the $D$ field is derived from a single particle model, and the filling sequence of zero energy LL ($N = 0,1$) additionally takes into account Coulomb interactions[31] (see also Supplementary Note 7). In **d–e,** colors indicate the type of $N\xi$, with solid lines for spin down and dashed lines for spin up. And, the bold lines indicate levels that are strongly influenced by the moiré potential in **a–c. f,** moiré-induced electronic energy spectra calculated for each valley. Note the Chern number $\Delta C$ of the moiré-induced isolated bands is different for each valley, leading to different valley-dependent broadening behaviors. **g,** $D$ field tunability of Chern insulator states in $N = 2$ Landau levels. **h,** Schematic of incompressible states in **g**. Each gap is color-coded and labeled $(t, s)$. Only t values are shown for IQHE and FQHE. Each color scale is truncated at the end value of its respective colorbar.

**Fig. 4 | Correlated insulating states. a,** Longitudinal resistance, as a function of normalized carrier density and magnetic flux, measured at $T = 30$ mK with $D = 91$ mV nm$^{-1}$. Blue arrows indicate insulating states at $n/n_0 = $ -1, -2 respectively. The color scale is adjusted to avoid saturations at $t$=0 features. **b,** Temperature dependence of longitudinal resistance of (0,-2) state at $\phi/\phi_0$=0.39 (indicated by a white arrow in **a**). The gap was estimated by fitting the Arrhenius formula $R_{xx} = R_0 \exp(\Delta/2 \mathrm{k_B} T)$.

**Fig. 5 | Interaction driven fractional states. a,** A zoomed-in area of **Fig. 2a** (see also **Supplementary Figure 1f**). The color scale is truncated at 15 kΩ. **b,** The Wannier diagram to denote observed states in **a**. Each incompressible state line is labeled with the same color as in **Fig. 2b**, and each colored region represents the single-particle Chern bands with Chern numbers $\Delta t$ obtained by evaluating the $t$-difference between the neighboring single-particle Chern gaps. The patterns of the Chern bands are similar for each of the $N$=0 (first and third from left) and $N$=1 (second and fourth) ZLLs. **c,** Line cuts along the $\phi/\phi_0 = 0.393$ line (dashed horizontal line in **a** and **b**). The blue graph on the left axis shows the Hall resistance in units of $h/e^2$ with dashed lines representing $1/t$. The orange graph on the right axis shows the longitudinal resistance. Each incompressible state is marked with the same colored arrow as in **b** and shows quantized plateau in Hall resistance and dip in longitudinal resistance. **d,** Temperature dependence of longitudinal resistance along the $\phi/\phi_0 = 0.389$ line. Each of the incompressible states exhibited thermal activation behavior, with resistance tending to increase with increasing temperature (see also **Supplementary Figure 6**). **e–h,** Temperature dependent longitudinal resistance of fractional states denoted in **d**. Gaps were estimated by fitting the Arrhenius formula with edge state $R_{xx} = R_0 \exp(-\Delta/2k_B T)$.

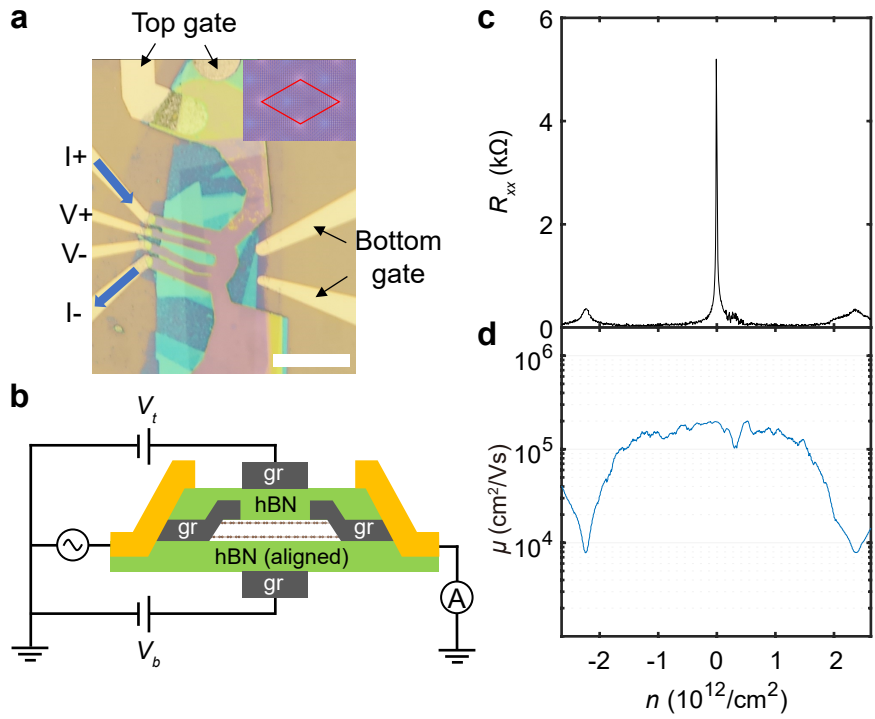

**a** Top gate

I+
V+
V-
I-

Bottom gate

**b**

$V_t$

gr
hBN
gr          gr
hBN (aligned)
gr

$V_b$

**c**

$R_{xx}$ (k$\Omega$)

0

**d**

$\mu$ (cm$^2$/Vs)

$10^6$

$10^5$

$10^4$

$n$ ($10^{12}$/cm$^2$)

-2    -1    0    1    2

**a**

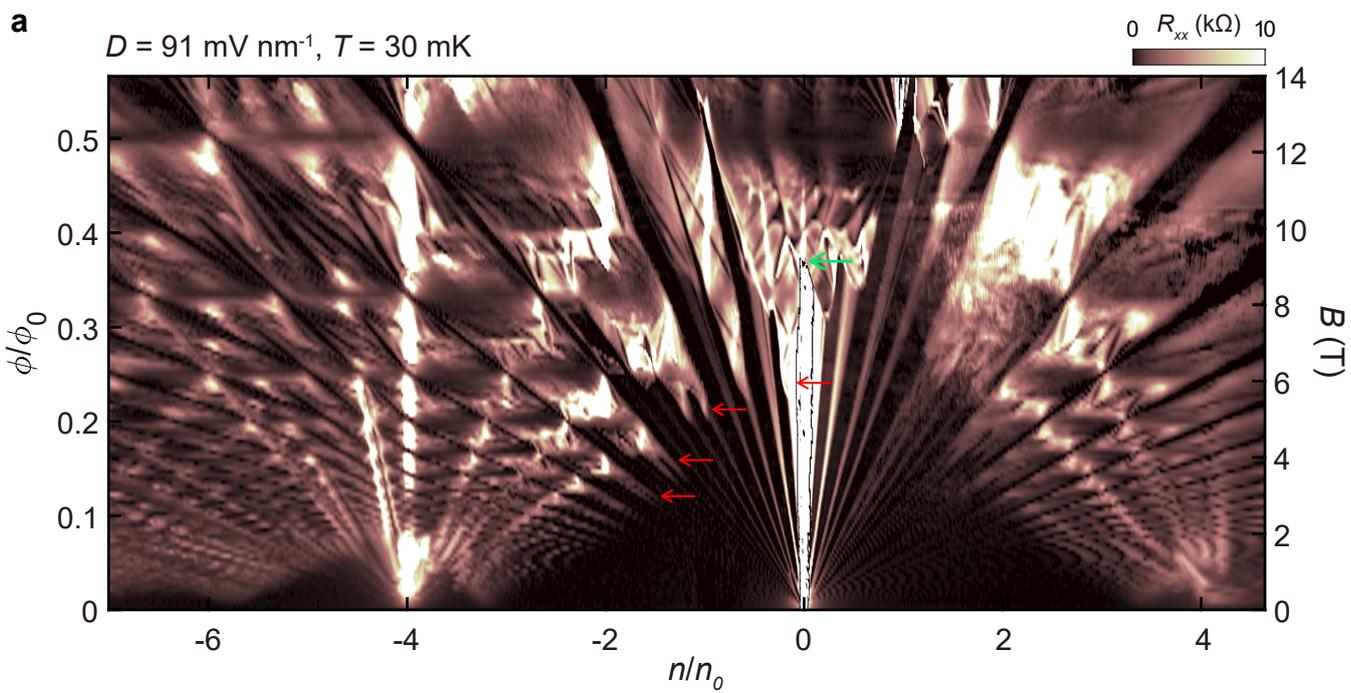

$D$ = 91 mV nm$^{-1}$, $T$ = 30 mK

$R_{xx}$ (kΩ)

0 — 10

$\phi/\phi_0$

$B$ (T)

$n/n_0$

**b**

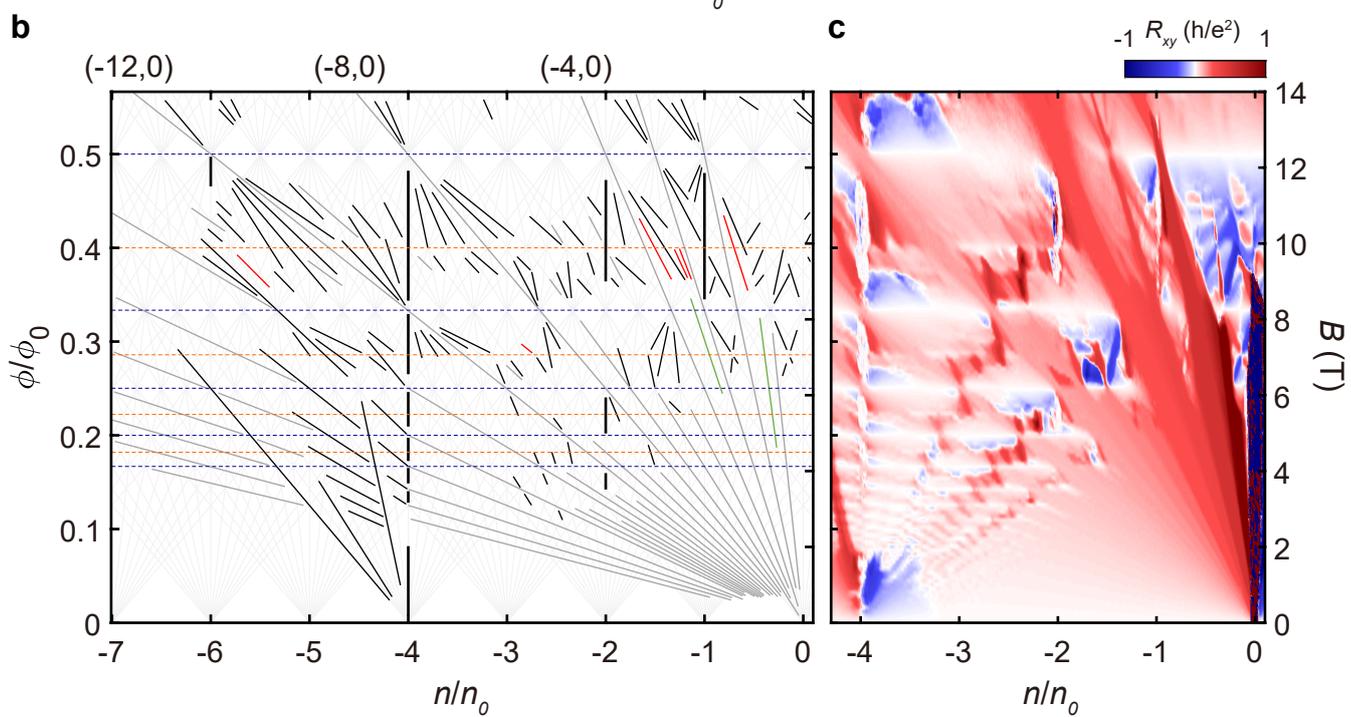

(-12,0)  (-8,0)  (-4,0)

$\phi/\phi_0$

$n/n_0$

**c**

-1 $R_{xy}$ (h/e$^2$) 1

$B$ (T)

$n/n_0$

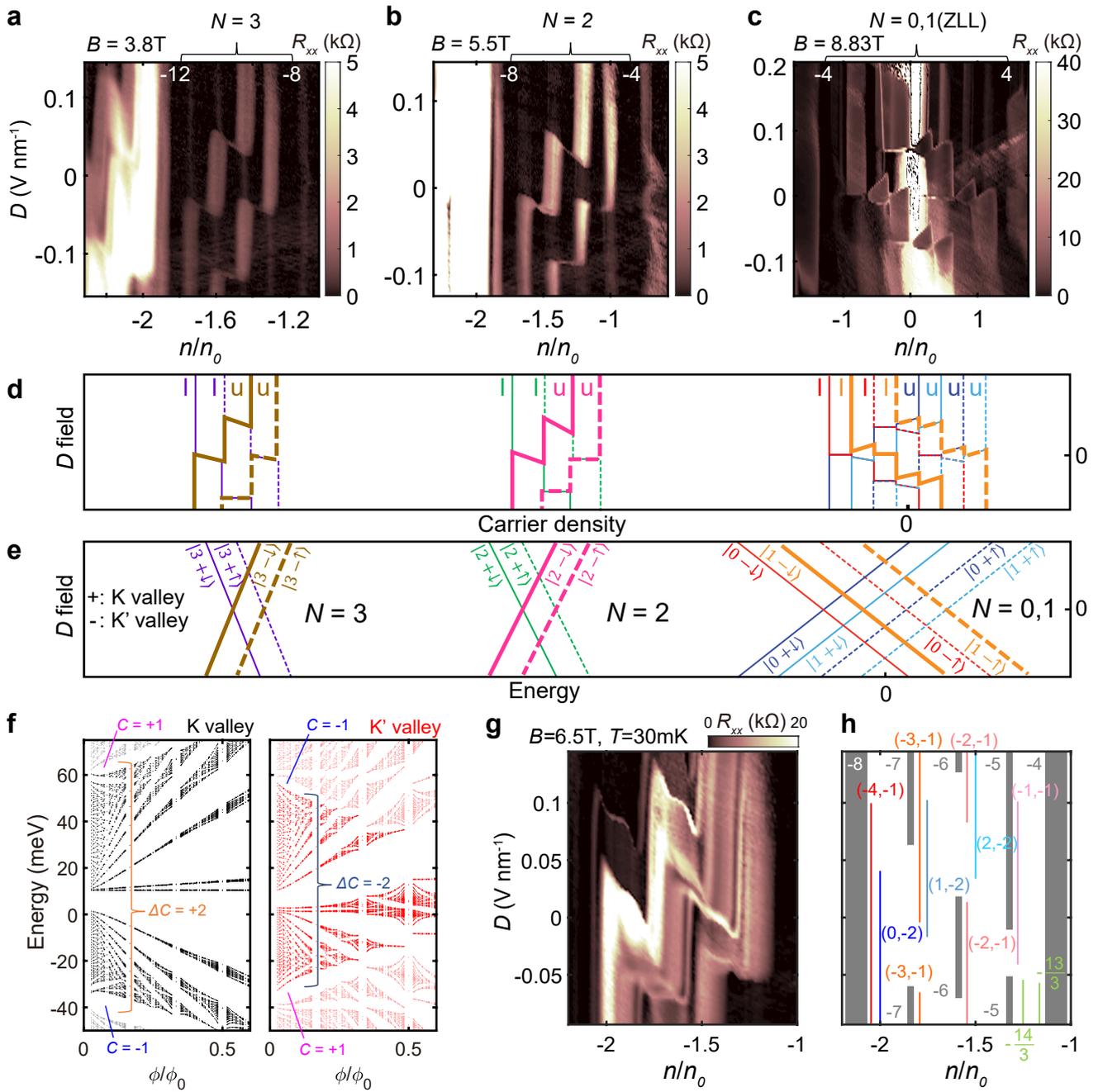

**a** B = 3.8T, N = 3, −12, −8, $R_{xx}$ (kΩ)

**b** B = 5.5T, N = 2, −8, −4, $R_{xx}$ (kΩ)

**c** B = 8.83T, N = 0,1(ZLL), −4, 4, $R_{xx}$ (kΩ)

**d** D field, Carrier density

l u u | l u u | l l l u u u

**e** D field, Energy

+: K valley, −: K' valley

$|3,+\rangle$ $|3,+\rangle$ $|3,-\rangle$ $|3,-\rangle$ | $N = 3$
$|2,+\rangle$ $|2,+\rangle$ $|2,-\rangle$ $|2,-\rangle$ | $N = 2$
$|0,-\rangle$ $|1,-\rangle$ $|0,+\rangle$ $|1,+\rangle$ $|1,-\rangle$ $|1,+\rangle$ $|0,+\rangle$ $|0,-\rangle$ | $N = 0,1$

**f** K valley, $C = +1$, $\Delta C = +2$, $C = -1$; K' valley, $C = -1$, $\Delta C = -2$, $C = +1$; Energy (meV), $\phi/\phi_0$

**g** B=6.5T, T=30mK, $0$ $R_{xx}$ (kΩ) $20$, $n/n_0$, D (V nm$^{-1}$)

**h** (−3,−1) (−2,−1), −8 −7 −6 −5 −4, (−4,−1) (−1,−1), (2,−2), (1,−2), (−2,−1), (0,−2), (−3,−1), −7 −6 −5, $\frac{14}{3}$ $\frac{13}{3}$, $n/n_0$

**a** $D = 91$ mV nm$^{-1}$, $T = 30$ mK

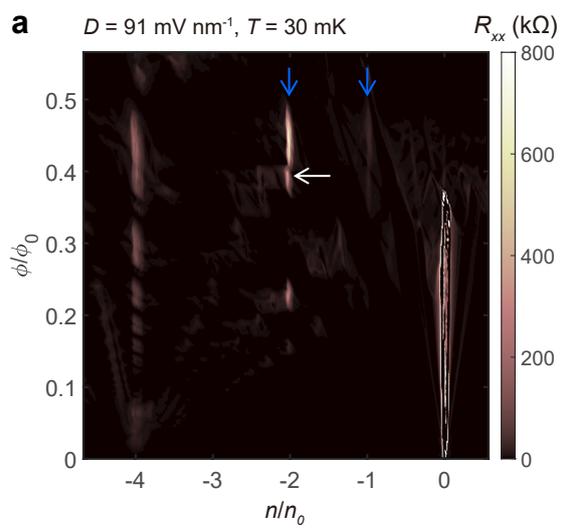

**b** $(t,s) = (0,-2)$   $\phi/\phi_0 = 0.39$

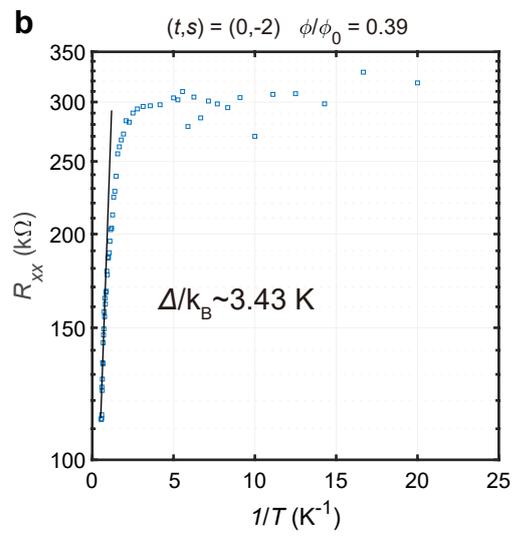

$\Delta/k_B \sim 3.43$ K

**a** $D = 91$ mV nm$^{-1}$, $T = 30$ mK

**b** (-5,1/2) (-5,2/3) (-4,1/3) (-3,1/2)

**c** $\phi/\phi_0 = 0.393$ ($B = 9.7$ T)

**e** (-5,1/2) $\Delta/k_B \sim 1.6$ K

**f** (-5,2/3) $\Delta/k_B \sim 110$ mK

**g** (-4,1/3) $\Delta/k_B \sim 250$ mK

**h** (-3,1/2) $\Delta/k_B \sim 1.9$ K

**d** (-5,1/2) (-4,1/3) (-5,2/3) (-3,1/2)

## Supplementary Information:

**Interplay of valley, layer and band topology towards interacting quantum phases in moiré bilayer graphene**

Yungi Jeong, Hangyeol Park, Taeho Kim, K. Watanabe, T. Taniguchi, Jeil Jung, and Joonho Jang

Correspondence to: joonho.jang@snu.ac.kr

**Supplementary Note 1. Additional measurement data**

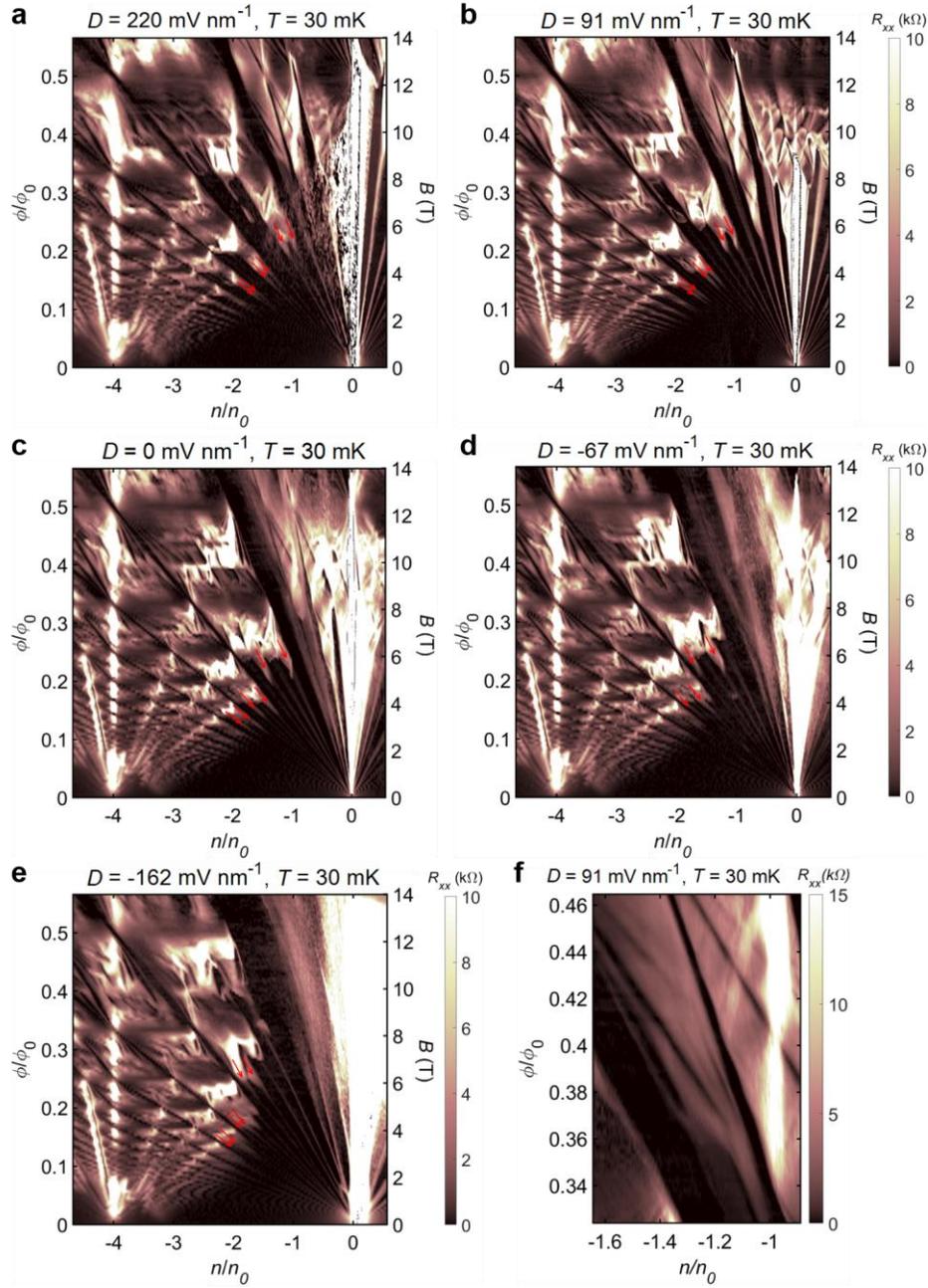

**Supplementary Figure 1 | Landau fan diagram with various *D* fields.** Longitudinal resistance versus carrier density and magnetic field at **a,** $D = 220\ mV\ nm^{-1}$, **b,** $D = 91\ mV\ nm^{-1}$ (higher resolution Fig.). **c,** $D = 0$, **d,** $D = -67\ mV\ nm^{-1}$, and **e,** $D = -162\ mV\ nm^{-1}$. Red arrows indicate metallic states that strongly feel the moiré potential within each $N \geq 2$ Landau level. As the *D* field is reduced from **b** to **e**, it can be seen that the positions indicated by the arrows move as shown in Fig. 3 of the main text. **f,** A more closely zoomed in image of Fig. 5a. Each color scale is truncated at the end value of its respective colorbar.

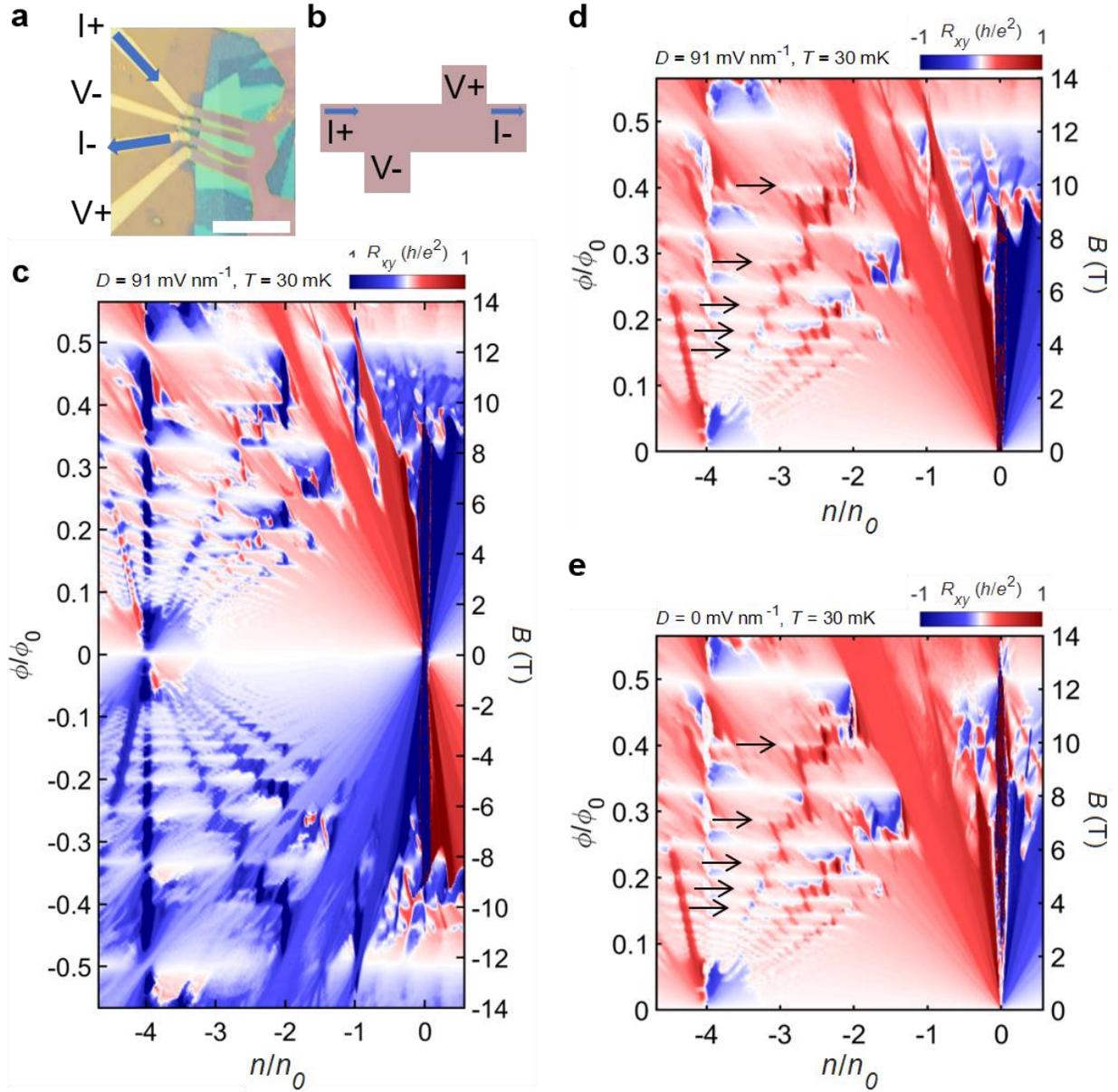

**Supplementary Figure 2 | Hall measurement a,** Hall measurement configuration. Current flows from I+ electrode to I- electrode and voltage (V+)-(V-) was measured. Scale bar, 15 μm. **b,** Schematic of the measurement configuration. Unfolding the horseshoe-shaped sample, it can be seen that **b** is topologically equivalent to **a**. And from this, we can expect a negative residual $R_{xx}$ component in the measurement. Especially, this effect is strong for vertical features at $n/n_0$ = -1,- 2,and -4 because these are not expected to have a chiral edge state. **c,** Hall resistance as a function of normalized carrier density and magnetic flux at $D = 91\ mV\ nm^{-1}$. **d–e,** Plot of $(R_{xy}(n, B) - R_{xy}(n, -B))/2$. at **d,** $D = 91\ mV\ nm^{-1}$ and **e,** $D$=0. The residual $R_{xx}$ component was removed by subtracting the measured value at the opposite magnetic field. This makes blue vertical lines have values close to zero as expected ($t = 0$). And each incompressible state has a quantized Hall resistance depending on the value of t, with a corresponding slope in the fan diagram. Along the

$\phi/\phi_0 = 1/q$ lines, the first order Brown–Zak oscillation occurs and the sign of $R_{xy}$ is reversed as the effective magnetic field felt by the BZ quasiparticle is reversed. In addition, horizontal lines with $R_{xy}$ close to zero are observed at the higher order BZ oscillation ($\phi/\phi_0 = 2/q$) lines (Black arrows in **d**,**e**).

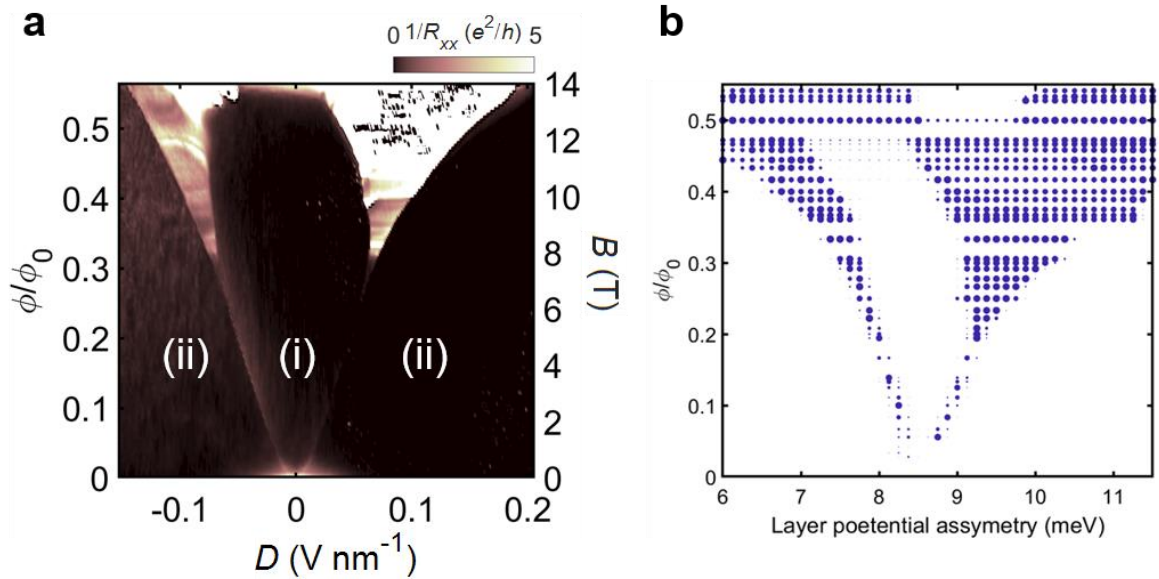

**Supplementary Figure 3 | Displacement field dependent CNP states a,** Reciprocal of Longitudinal conductance in units of $e^2/h$ as a function of applied perpendicular electric displacement field and magnetic field at $n = 0$ (CNP) and $T = 30$ mK. Labels (i) and (ii) indicate the two different insulating $\nu = 0$ phases; (i) canted antiferromagnet state and (ii) layer polarized insulator state known in pristine BBG.[1] There is a phase transition point between the two phases where the conductance increases. But, as the magnetic field increases, this phase transition point widens and a metallic region appears between (i) and (ii). **b,** A numerical simulation of gap closings (with the marker size inversely proportional to the gap size) near CNP for the same conditions is presented for comparison, showing a good agreement with the measurement. This gap closing is likely due to the overlapping of the broadened conduction and valence bands under the influence of the moiré potential.

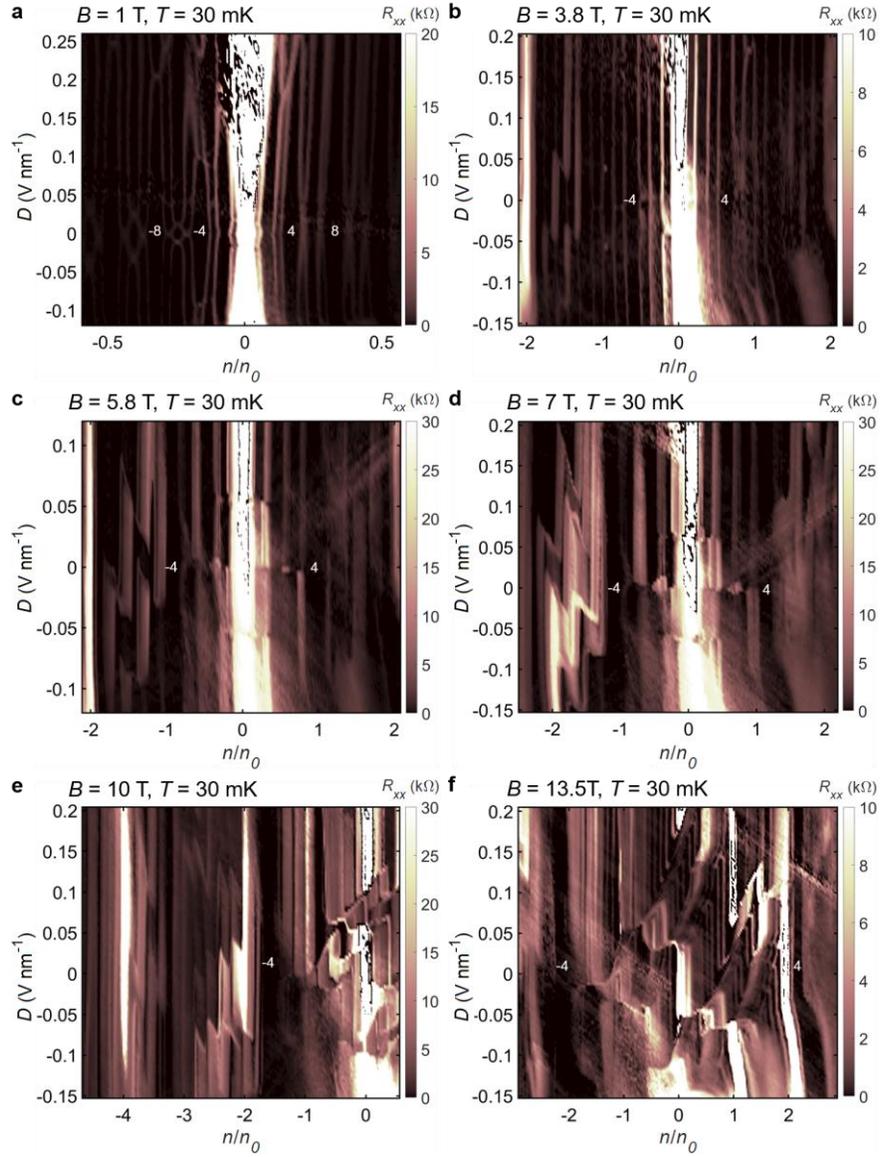

**Supplementary Figure 4 | $n - D$ sweep at various magnetic field.** Longitudinal resistance versus carrier density and $D$ field at **a,** $B = 1$ T, **b,** $B = 3.8$ T, **c,** $B = 5.8$ T, **d,** $B = 7$ T, **e,** $B = 10$ T, and **f,** $B = 13.5$ T. Numbers in each figure represent the Landau filling factor $\nu$. **a** shows that at the low magnetic field, there is little effect from the moiré potential and a single particle Landau level spectrum of intrinsic BBGs with broken spin–valley degeneracy appears. Around $D = 0$, intra-Landau level crossings for each N were observed, and for large $D$ field, inter-Landau level crossings between different $N$s also occurred.[2] **b** is the zoomed-out image of Fig. **3a**. Two of four $N = 3$ LL feels the moiré potential while $N \leq 2$ LLs are not yet affected by the moiré potential. In **c,d**, it is observed that the states with strongly feeling moiré potentials in the ZLL and $N = 2$ LL shift in opposite directions as D is varied. **e,f** shows that a large number of CIs undergo complex transitions, and the insulating gap at CNP closes in a certain $D$ field range. Each color scale is truncated at the end value of its respective colorbar.

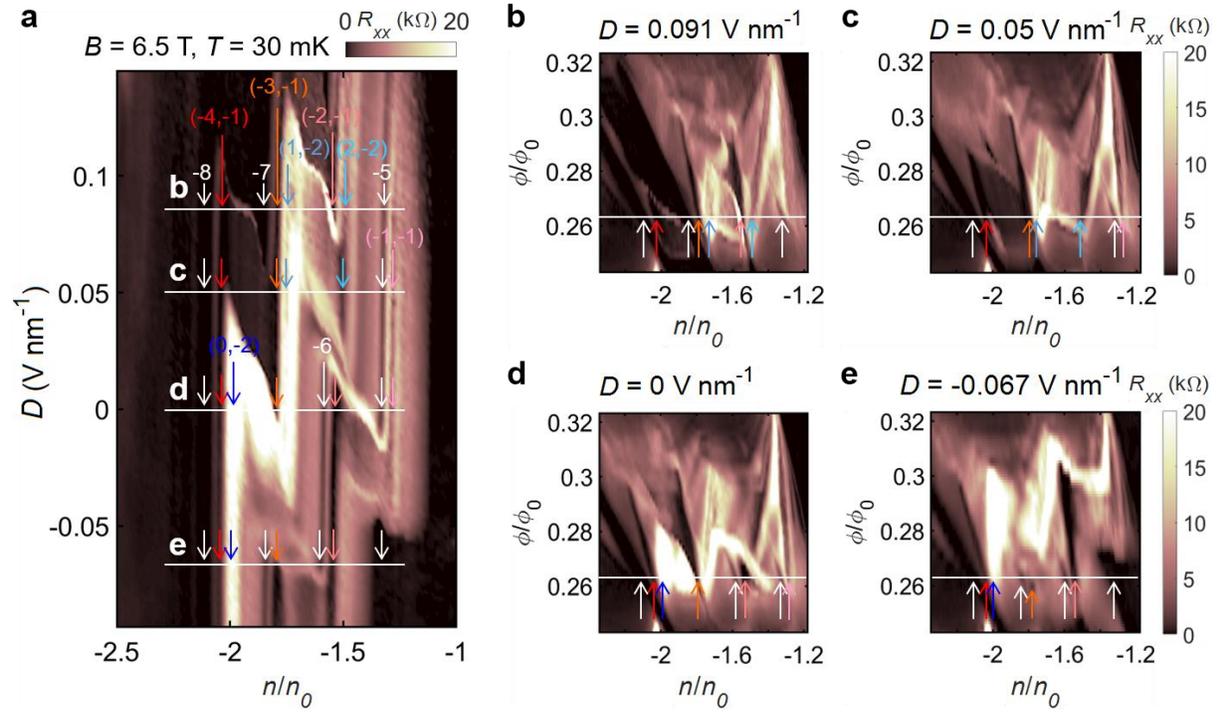

**Supplementary Figure 5 | Displacement field tuning of Chern insulator states. a,** Longitudinal resistance versus carrier density and displacement field at $B = 6.5$ T ($\phi/\phi_0 = 0.263$), $T = 30$ mK which is the same region as in **Fig.3f**. Each white horizontal line corresponds to a white horizontal line in **b–e** and each incompressible state$(t,s)$ is pointed by the same colored arrows as in **Fig.3g** ($s$ is omitted for IQHE states). **b–e,** Local Landau fan diagram of longitudinal resistance at different perpendicular displacement fields. Each state is represented by a line segment satisfying the Diophantine equation in the Landau fan diagram by the corresponding $(t,s)$. Each color scale is truncated at the end value of its respective colorbar.

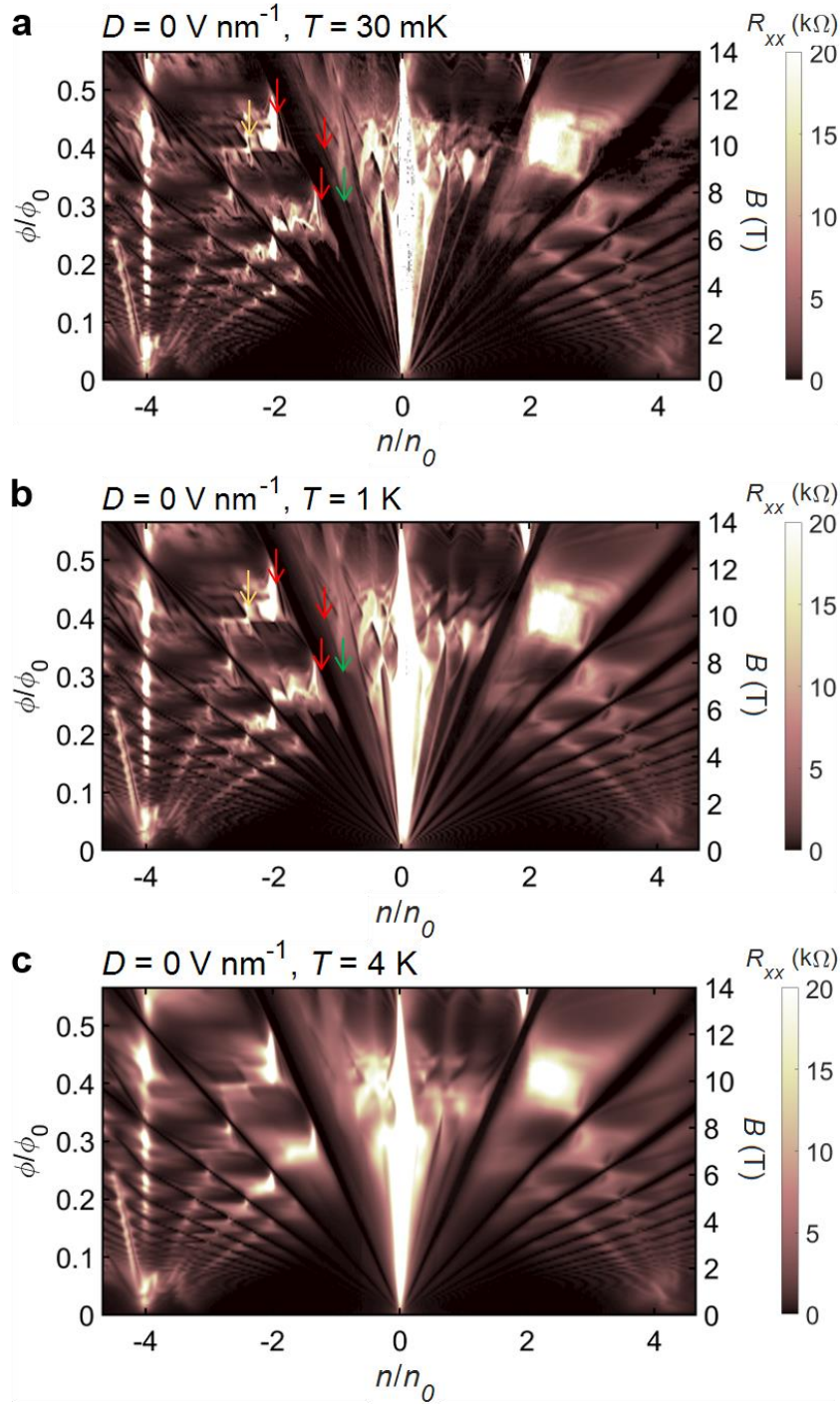

**Supplementary Figure 6 | Landau fan diagram with various temperatures at $D = 0$.** Longitudinal resistance versus carrier density and magnetic field at **a,** $T = 30$ mK, **b,** $T = 1$ K, and **c,** $T = 4$ K. Each color scale is truncated at the end value of its respective colorbar. At $T = 1$ K, the CI (yellow arrow), SBCI (red arrow), and FQHE (green arrow) states disappeared. And at $T = 4$ K, most of the CIs have melted away.

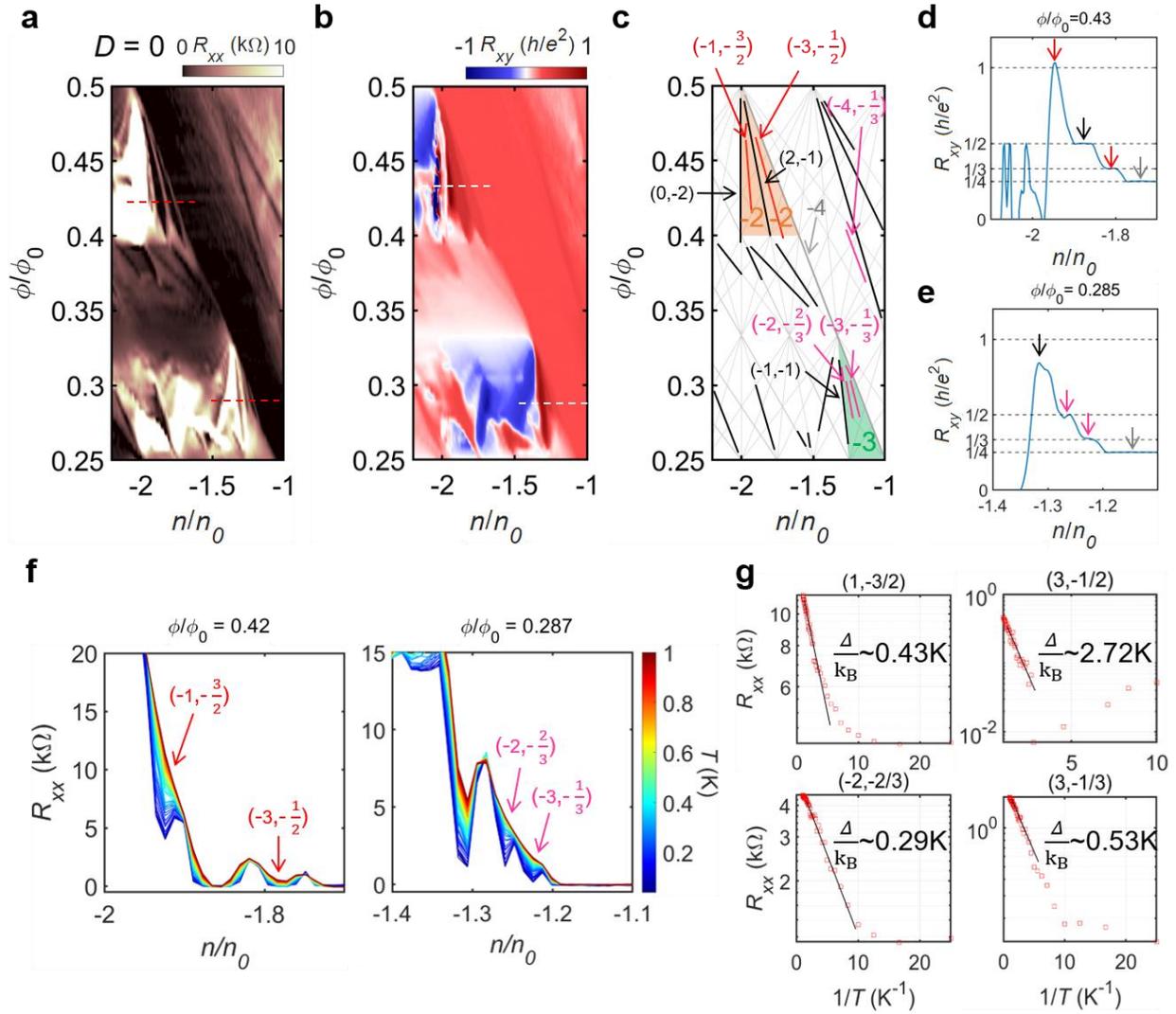

**Supplementary Figure 7 | Symmetry broken Chern insulator states in higher Landau level at zero displacement field. a,b** A zoomed in image of longitudinal(Hall) resistance Landau fan diagram at $D = 0$ (See **Supplementary Figure 1c** and **2e**). The color scale is truncated at 10 k$\Omega$ in **a**. **c,** Wannier diagram of the same area as **a, b**. The upper SBCI has $(t,s)$ = (-1,-3/2), (-3,-1/2), from left to right, and the lower SBCI has (-2,-2/3), (-3,-1/3), respectively. The upper SBCIs are half-filled with the band with $\Delta t$ = -2, and the lower SBCIs are 1/3 and 2/3 filled with the band with $\Delta t$ = -3, respectively. **d–e,** Line cuts along the upper (**d**), lower (**e**) dashed line in **b**. The colored arrows indicate the states corresponding to the same colored lines in **c**. Each of the incompressible states showed a well quantized Hall resistance except the CI (-1, -1) state in **e** which didn't reached $h/e^2$. This is probably due to a part of the $R_{xx}$ component (bright metallic part surrounding the gap) was not removed during the removal process, or the measurement speed didn't allow the value to saturate as the t value flipped from 2 to -1. **f,** Temperature dependence of longitudinal resistance along the dashed lines in **a**. **g,** Arrhenius plots of identified SBCIs.

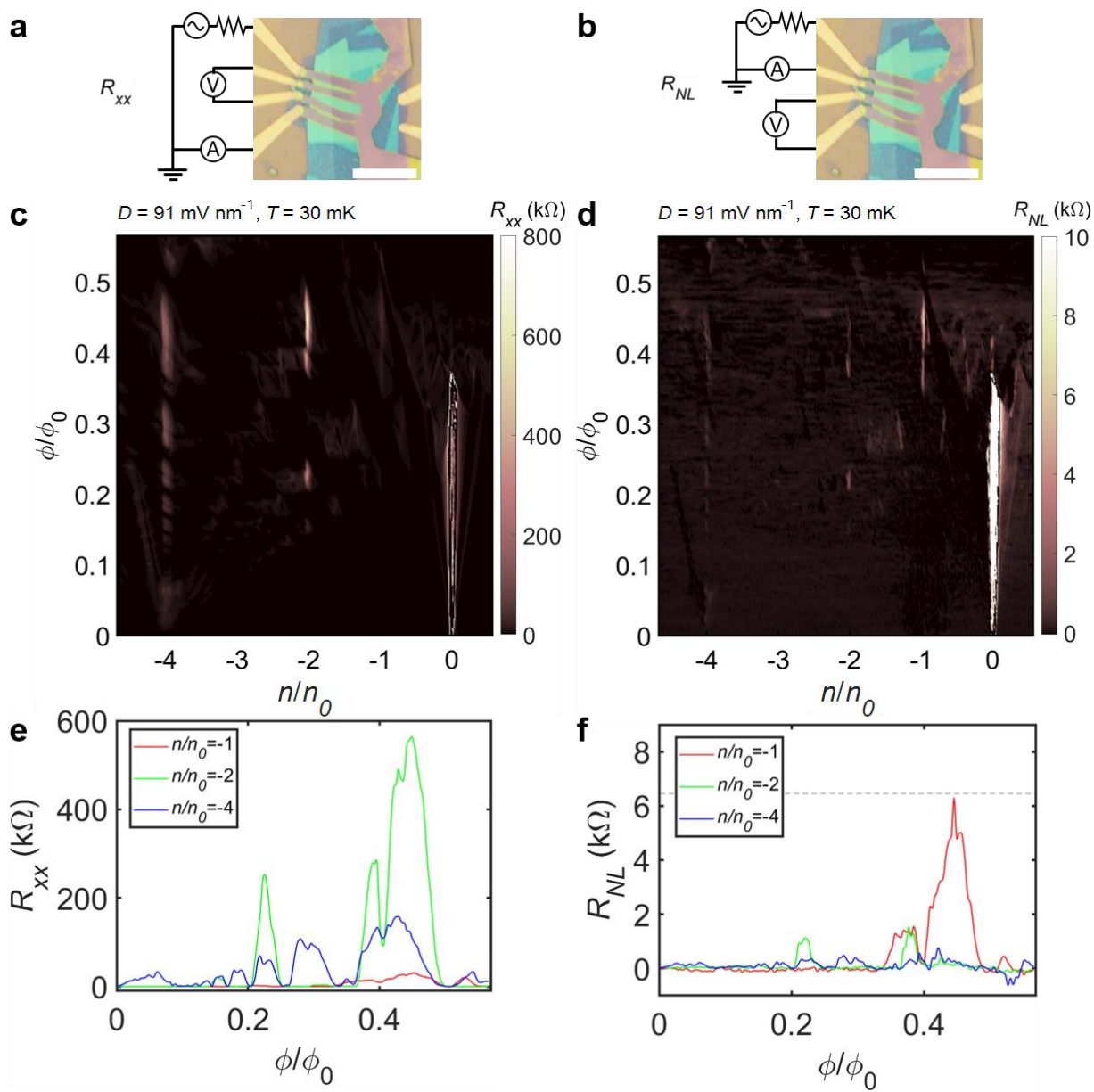

**Supplementary Figure 8 | Nonlocal measurement of helical edge states at t=0 insulators. a–b,** Measurement configuration of longitudinal ($R_{xx}$, **a**) and non-local ($R_{NL}$, **b**) resistance. Scale bar, 15 µm. **c–d,** Longitudinal (**c**) and non-local (**d**) resistance, as a function of normalized carrier density and magnetic flux, measured at $T = 30$ mK with $D = 91 mV\ nm^{-1}$. **e–f,** Line cuts of longitudinal (**e**) and non-local (**f**) resistance along the magnetic field direction at $n/n_0 = $ -1 (red), -2 (green), and -4 (blue). Dashed horizontal line in **f** represents $h/4e^2$.

**Supplementary Note 2. Device fabrication**

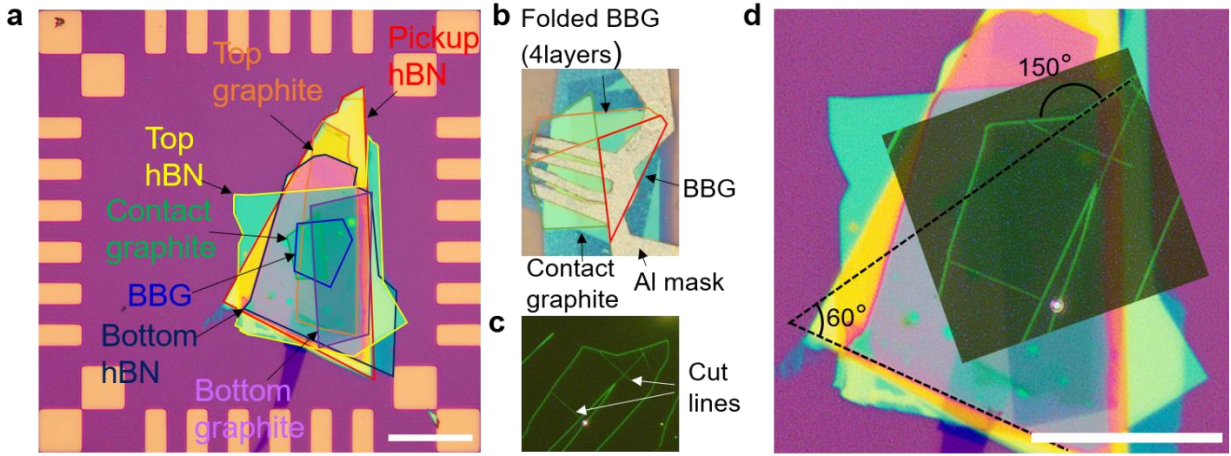

**Supplementary Figure 9 | BBG/hBN aligned heterostructure a,** Optical microscope image of the heterostructure stack. Each component is outlined in a different color and indicated by an arrow. **b,** Zoomed-in image of the stack during the reactive ion etching(RIE) process. Since graphene and hBN have a very high selectivity for $CF_4$ etching, the contrast between the graphene-protected and non-graphene-protected regions during RIE is dramatic, making it easy to see the alignment of the BBG and hBN. **c,** Dark field image of the BBG flake used in fabrication. A long BBG flake was cut by the scanning probe based local anodic oxidation method.[3] **d,** Overlapped image of **a** and **c**. **b** was used for alignment. It can be seen that one crystal axis of BBG is aligned at 60 degrees to the crystal axis of the bottom hBN. Scale bar, 30 μm.

**Supplementary Note 3. Transport data axis normalization and superlattice size estimation.**

The x-axis of the raw data in the Landau fan diagram is the gate voltage proportional to the carrier density, and the y-axis is the perpendicular magnetic field. To convert this to superlattice filling factor and normalized flux density, top gate and bottom gate sweep and Hall measurement were performed at low field. The carrier density n and vertical electric displacement field $D$ can be expressed as follows using the top gate voltage $V_t$ and bottom gate voltage $V_b$.

$$n = \frac{c_t V_t + c_b V_b}{e} = \frac{c}{e}\{(\cos\theta\,)V_t + (\sin\theta\,)V_b\} \equiv \frac{c}{e}V_x$$

$$D = \frac{-c_t V_t + c_b V_b}{2\epsilon_0} = \frac{c}{2\epsilon_0}\{-(\cos\theta\,)V_t + (\sin\theta\,)V_b\} \equiv \frac{c}{2\epsilon_0}V_y$$

$$c = \sqrt{c_t{}^2 + c_b{}^2}, \qquad \theta = \mathrm{atan}\left(\frac{c_b}{c_t}\right)$$

Where e is the unit charge, $\epsilon_0$ is the permittivity of vacuum, $c_t$ and $c_b$ are the capacitance per unit area between the sample and the top gate and bottom gate respectively. If we think of $(c_t\,, c_b)$ as a single two-dimensional vector and call the norm of this vector c and the polar angle $\theta$, we can define $V_x$, the voltage proportional to the carrier density, and $V_y$, the voltage proportional to the $D$ field, as above. Then, if we know $\theta$ and c, we can find the carrier density and $D$ field from the voltage. First, to measure $\theta$, a longitudinal resistance measurement was performed by sweeping the top gate voltage and bottom gate voltage in the magnetic field of 0.5T.

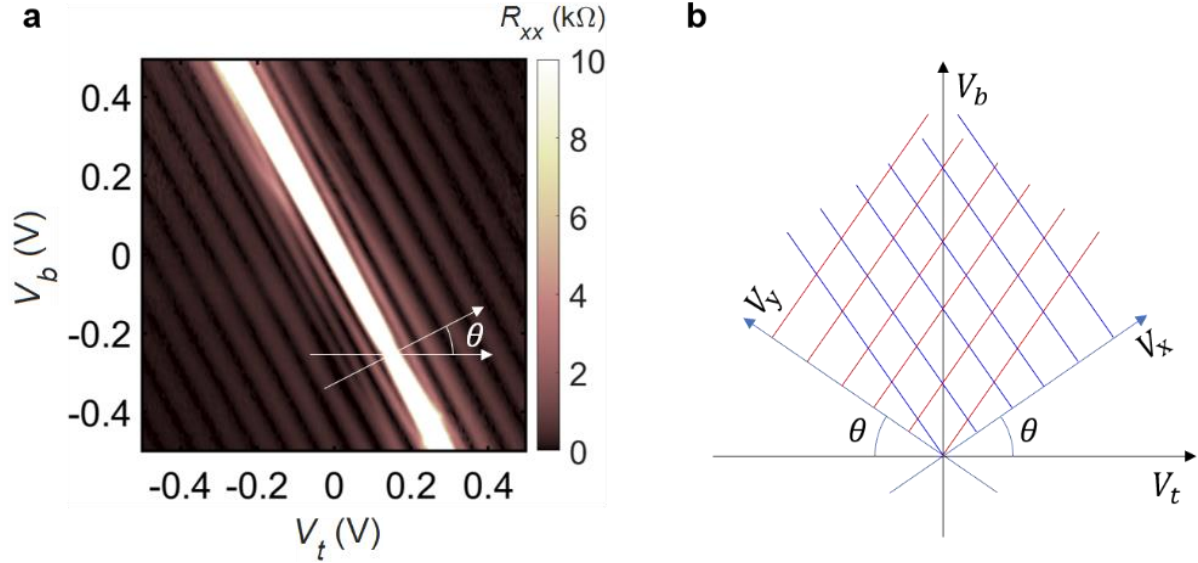

**Supplementary Figure 10 | Capacitance ratio determination. a,** Longitudinal resistance was measured by sweeping the top and bottom gate voltage at $B = 0.5$ T, $T = 30$ mK. Bright white diagonal line at the center is the CNP of BBG and each black line represents an integer quantum Hall state. By measuring the angle between $V_t$ direction and $\mathrm{grad}(n)$ direction, we can get the ratio between the top and bottom capacitance. $\theta$ was 0.505 rad in our device. **b,** Schematic diagram of axis conversion. Blue and red lines perpendicular to $V_x$ and $V_y$ are the equipotential line of each variable. $\theta$ is the angle defined the same as in **a**.

The darker colored regions with lower resistance are the integer quantum Hall states with chiral edge modes, and the lighter regions in between are CNP (brightest line) and the metallic states with partially filled Landau level extended states. The direction perpendicular to these diagonal lines is the direction of increasing carrier density, and by measuring the angle between this direction and the x-axis ($V_t$ axis), $\theta$ defined above can be obtained, and $\theta = 0.505$rad was measured. From this, it can be seen that $c_b = 0.5528c_t$. By fixing $V_y$, $D$ field can be fixed and only the carrier density can be tuned. And to measure c, low field Hall measurement was performed

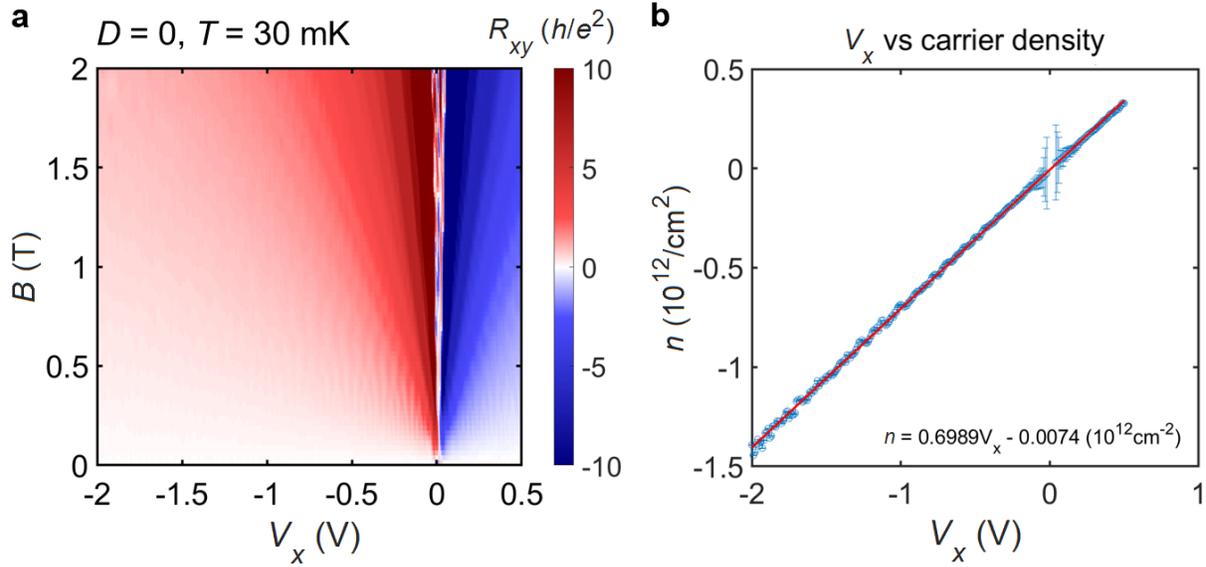

**Supplementary Figure 11 | Capacitance size determination. a,** Low field Hall resistance measurement. In the low carrier density region, quantum Hall is already visible below 1 T, but as the density increases, $R_{xy}$ shows a linear relationship as predicted by the Drude model. **b,** Measured $V_x$ dependence of the carrier density. From the Drude model, $R_{xy}(B) = -\frac{B}{ne}$. So the slope of $R_{xy}(B)$ is proportional to the reciprocal of the carrier density, which is calculated and plotted for each $V_x$. Error bars are proportional to the slope error of $R_{xy}(B)$ at each $V_x$. From the linear fit, we can see that $n = 0.6989V_x - 0.0074$ ($10^{12}\ cm^{-2}$), and using the fact that the coefficient of $V_x$ is $c/e$, we can see that $c = 1.1198 \times 10^{-15}\ F\ \mu m^{-2}$.

The Hall measurement was performed in the configuration shown in **Supplementary Figure 2a,** and since it is not an ideal Hall bar geometry, it measures a mixed value of $R_{xx}$ and $R_{xy}$, but at the Drude model level, $R_{xx}$ is independent of the magnetic field, so the slope can be used to determine the carrier density. From the measurement results, we determined the relationship between carrier density n and $V_x$.

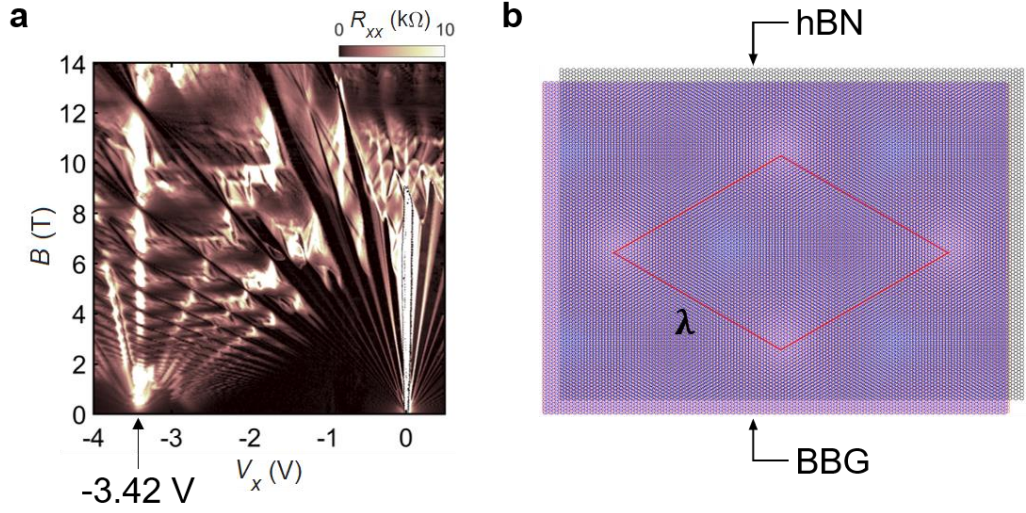

**Supplementary Figure 12 | Determination of superlattice lattice parameter. a,** Raw data of the Landau fan diagram. Longitudinal resistance was measured as $V_x$ and $B$ were swept. Superlattice band insulator was observed at $V_x = -3.42\ V$. This corresponds to the carrier density $n = 2.397 \times 10^{12}\ cm^{-2}$. **b,** moiré superlattice pattern of 0 degree aligned BBG/hBN heterostructure. The lattice constant mismatch of about 1.8% generates a triangular superlattice structure with the lattice constant $\lambda$.

From **Supplementary Figure 12**, we can see that a band insulator appears at $V_x = -3.42\ V$. This can occur when the moiré superlattice is filled with all four flavors of spin and valley. The superlattice parameter $\lambda$ can be calculated from the corresponding carrier density and the fact that the moiré superlattice is a triangular lattice.

$$n_0 = \left(\frac{\sqrt{3}}{2}\lambda^2\right)^{-1}$$

$$4n_0 = 2.397 \times 10^{12}\ cm^{-2} \ \Rightarrow\ \lambda = 13.88\ nm$$

From the magnitude of the superlattice parameter, we can see that our sample has BBG and hBN aligned at almost 0 deg.

Alternatively, the superlattice parameter can be obtained independently by observing the Brown–Zak oscillation. From **Fig. 2** and **Supplementary Figure 2**, $\phi/\phi_0 = 1/q$ line is equivalent to $B = (24.7/q)\ T$ line, so we can estimate the superlattice unit cell area by $A \times 24.7T = \phi_0 = h/e$, $A = \frac{\sqrt{3}}{2}\lambda^2$. The resulting $\lambda$ is 13.90 nm which is almost the same value obtained by the above method.

**Supplementary Note 4. Numerical simulation of Hofstadter energy spectra and Wannier plots of the BBG/hBN superlattice with zero-degree alignment**

After electronic energy eigenvalues are obtained by numerically diagonalizing at various perpendicular magnetic fields, we transform it to the Wannier plot as a function of density and magnetic field as in **Supplementary Figure 13**. It shows features that can be directly associated with our measured data. First of all, in the low magnetic fields, the LL band broadening due to the moiré potential is selectively strong for a certain valley (see also **Supplementary Figure 15**). As discussed in the main text, it is counterintuitive that the valley degrees of freedom determine the strength of the superlattice effect, not the layer degrees of freedom.

The multiple plots by varying the interlayer potential difference in **Supplementary Figure 14** — where we define $D$ field = (interlayer potential difference) / (interlayer distance) — shows the pronounced features that strongly depend on the $D$ field. We reproduce the insulating gaps at the full fillings, the characteristic spectra of the ZLL and the valley degrees of freedom's dependence on the external vertical electric displacement field, etc. Importantly, the vertical insulating phases at $n/n_0 = -2$ and $-4$ measured in **Fig. 4** of the main text are either absent or having energy gaps just similar to other incompressible Chern phases in the calculation. This thus supports our conclusion that the strong insulating phase at $n/n_0 = 2$ has an enhanced energy gap due to the electron correlation, and it is likely the consequence of the narrow bandwidth of the isolated valence band (see the calculated spectra in **Supplementary Figure 13a–d**), as discussed in the main text.

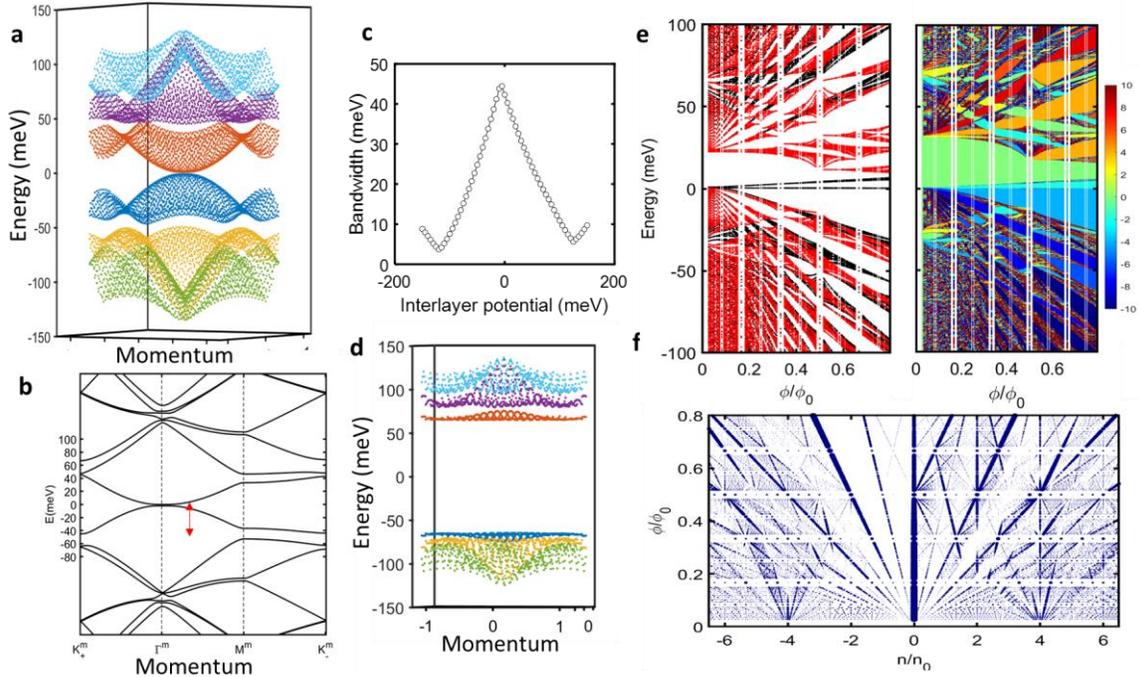

**Supplementary Figure 13 | Hofstadter spectra and Wannier plots as a function of the interlayer potential difference. a–b**, Zero magnetic field spectra as a function of momentum in the moiré Brillouin zone. **c,** The bandwidth calculated for the first valence band as a function of the interlayer potential. Interestingly, the band even becomes nearly flat in a strong $D$ field, while not accessible in our current sample. **d**, The energy spectra near the minimum bandwidth in **c**. **e**, Magnetic field induced Hofstadter spectra at interlayer potential $U_{layer} = 0$. Two distinct valleys are indicated in black (K) and red (K') colors, and the Chern numbers for energy gaps are indicated with colors. **f**, Wannier diagram calculated from **e**. The size of the gaps are represented with the size of a marker, and only energy gaps larger than 0.1 meV are plotted. Note that spin degrees of freedom are assumed to be degenerate.

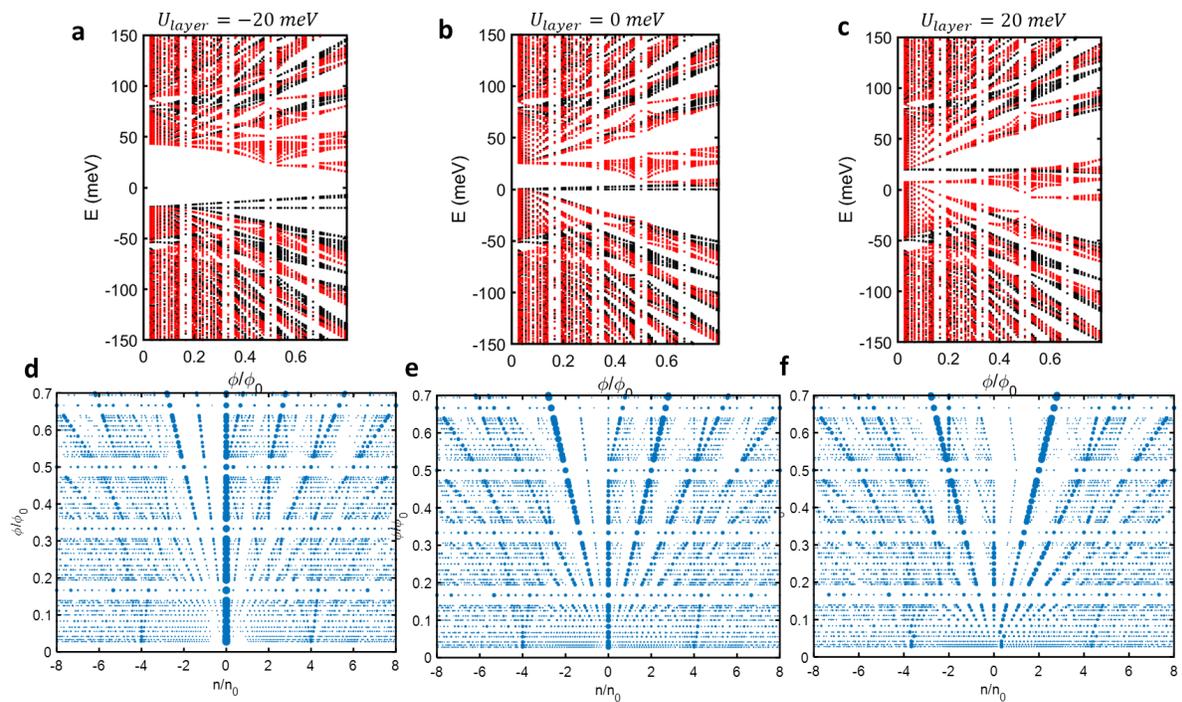

**Supplementary Figure 14 | Electron energy spectra and Wannier plots for three different interlayer potential differences. a–c**, In the spectra, the valleys are plotted in black (K) and red (K') colors separately. **d–f**, The Wannier plots only show energy gaps whose values are larger than 0.1 meV.

**Supplementary Note 5. Numerical simulation of the correlation of valley, layer and Landau level index of the BBG/hBN superlattice**

Based on the continuum model simulation of the previous section, we further calculated the valley-dependent energy spectra as a function of the interlayer potential at a fixed magnetic field. First of all, due to the dominant cyclotron energy scale, the LLs are sharply defined in the case of intrinsic BBG, and the moiré-broadened LLs are still well-resolved for the BBG with aligned hBN. The valley-dependent bands are plotted separately in blue (K) and red (K') colors. From the intra-LL energy crossings of valley-dependent bands (**Supplementary Figure 15b**), we clearly see a peculiar correlation between the valleys, layers and LL indices to support the conclusion drawn based on **Equation (1)** in the main text. We further show the filling sequence of bands based on the energy spectra and, in **Supplementary Figure 15d**, obtain qualitatively similar to our transport measurement to **Fig. 3**. Most strikingly to us, the simulation results clearly confirm that the effects of the moiré superlattice from the hBN interface are selectively felt by one of the valleys (namely K'), and we attribute this behavior to be the cause for the striking "fork" features (i.e. LL subbands with high longitudinal resistance in the Landau fan) discussed in the main text. In the following section, we further discuss the possible topological origin of the valley-selective effects based on the simulation.

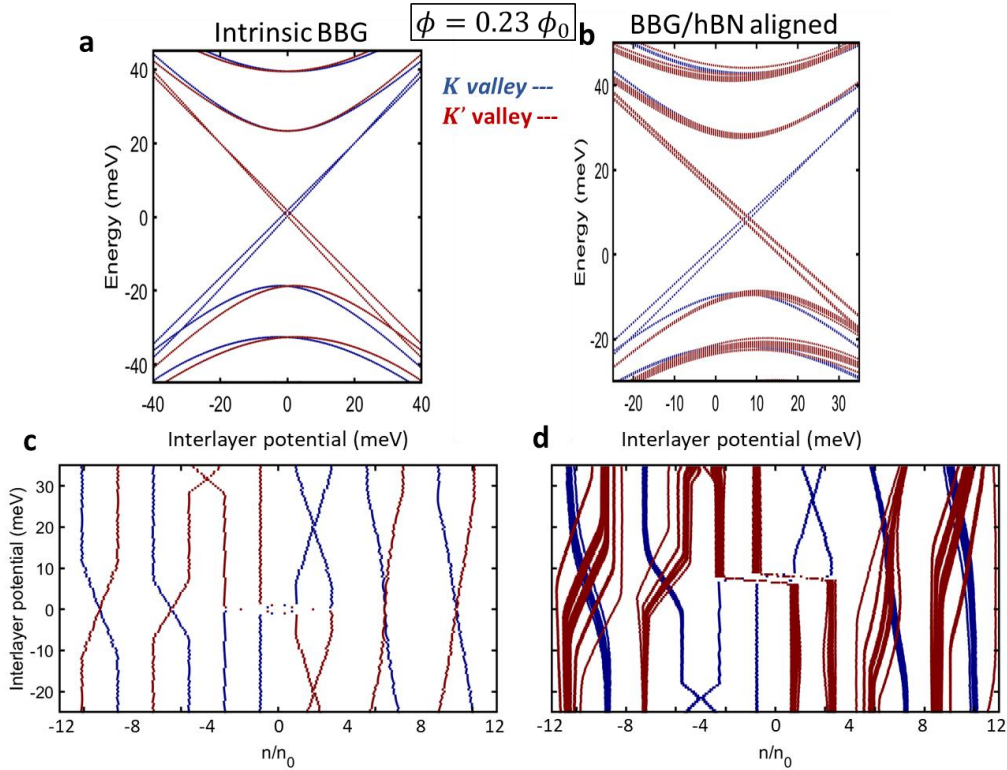

**Supplementary Figure 15 | Electronic energy spectra and filling sequences calculated based on a continuum model**, as explained in the text, for an intrinsic BBG (**a** and **c**) and an BBG aligned with hBN (**b** and **d**), as a function of the interlayer potential difference and density at $\phi = 0.23\phi_0$ .

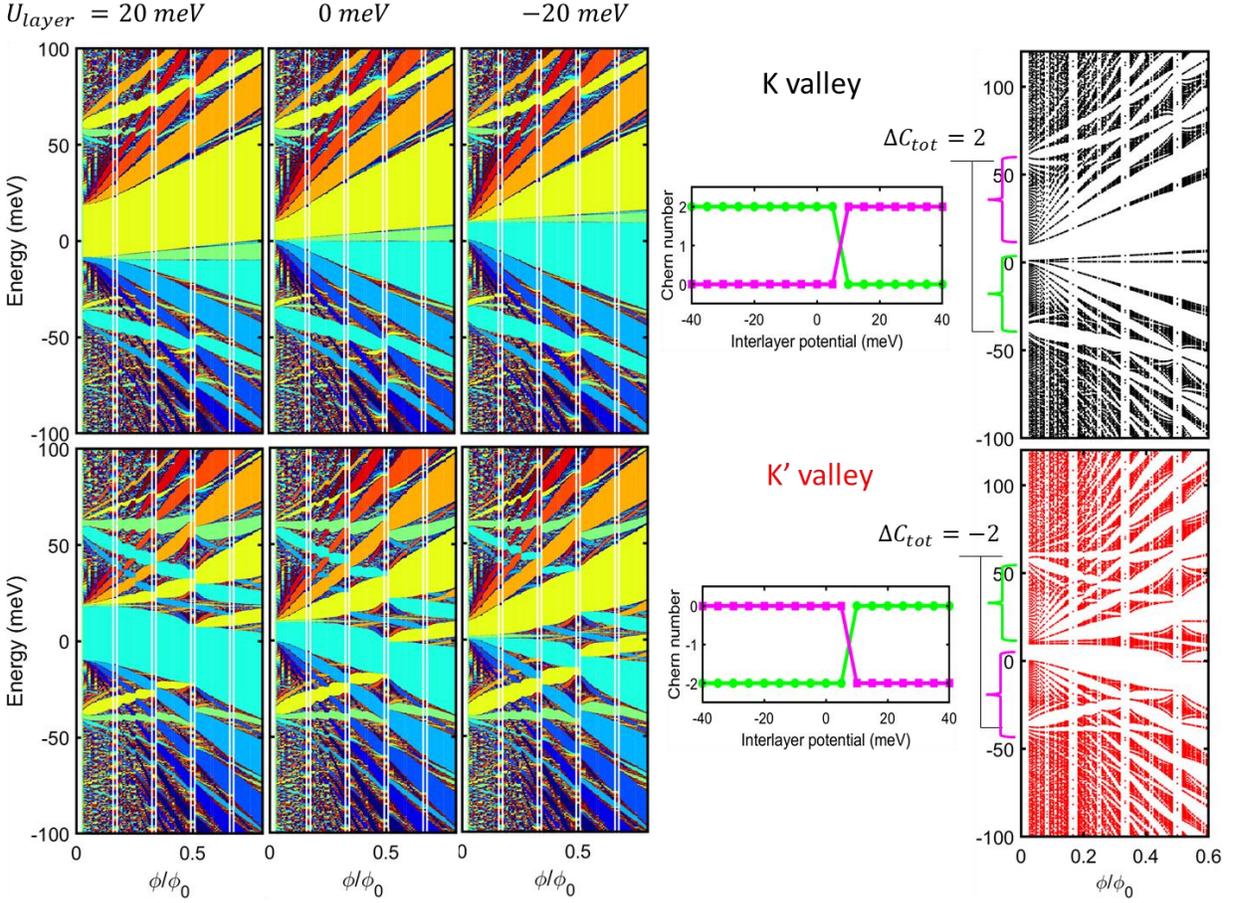

**Supplementary Figure 16 | Magneto-spectra for K and K' valleys.** At various interlayer potentials, Chern numbers of energy gaps are determined by evaluating $\partial n/\partial \phi$. The Chern numbers of the isolated bands are also evaluated as difference of Chern numbers below and above the bands. In the mid panel, the Chern number of isolated bands at conduction and valence bands are plotted with green and magenta colors, respectively. The total Chern numbers of the two moiré-induced isolated bands ($-4 < n/n_0 < 4$) don't change when the interlayer potential is varied, and the value is clearly valley dependent. In the Hofstadter spectra, we used black (K) and red (K') colors to indicate valleys.

**Supplementary Note 6. Discussion on the valley-selective moiré effect and valley-dependent Chern numbers of moiré-induced isolated bands**

The valley selective response of the system to the moiré patterns in the presence of an external magnetic field is likely due to the valley selective partial cancellation and reinforcement of the vector potential generated by the magnetic field itself and the virtual strains affecting the system. The vector potential due to a magnetic field breaks time reversal symmetry and leads to asymmetric response of the bands at the K and K' valleys. The virtual strains arise due to second order perturbative hopping effects of the graphene electrons to the hBN atoms and back, leading to time reversal symmetric vector fields for each valley. The pseudomagnetic fields can also result from local strain fields that introduce unequal hopping terms of the electrons at a given lattice site to the neighboring surrounding atoms. These real or virtual strain profiles lead to strong pseudomagnetic fields ranging from a few to a few tens of Teslas at each point in space that have opposing signs for electrons of each valley. However, a real magnetic field has a definitive sign for the electrons in both K and K' valleys.

In the simulation, we find the Chern number of the moiré-induced isolated bands ($-4 < n/n_0 < 4$) have the exact opposite values for each valley (**Supplementary Figure 16**), similar to the observation in ref[4–6]. We attribute this observation is closely related to the observed valley-selective moiré effect. Apparently in the simulation, the Chern numbers for the insulating gap at the conduction band ($n/n_0 = 4$) and for the gap at the valence band ($n/n_0 = -4$) change the sign upon changing the valley quantum number, independent of $D$ field. This in turn leads to the disparate Hofstadter spectra: The effective magnetic moments from the angular momentum of the bands result in the phenomena that the bands of a valley with the anti-parallel band moments and the magnetic field experience more frequent intersections between Hosftadter gaps with opposite Chern values, leading to stronger moiré effects. Moreover, interestingly, based on the simulation, the calculated valley Chern numbers of the induced gaps ($n/n_0 = 4$ and $-4$) in the valence band seem to suggest that they are quantum valley Hall insulators. The value of the longitudinal resistance for the insulators at $n/n_0 = -4$ at the base temperature appears significantly lower than the expected value from the calculated gap.

This valley selective moiré effect is also observed in Landau levels beyond the lowest ones near the band edges at charge neutrality, namely those that are most directly sensitive to the signs of the applied electric fields. Especially in the limit of small magnetic fields where the Landau levels are closely spaced in energy, it is expected that the Landau level mixing facilitates the propagation of the sublattice/layer/valley-selective moiré effect to higher Landau levels especially when the ground-state solutions are of spin or charge density wave type that favors accumulation of spin-resolved charge in a given sublattice.

**Supplementary Note 7. Filling sequence of zeroth Landau level in BBG with interaction**

If we draw the energy spectrum (or the filling sequence) of the zero Landau level (ZLL) based on the single particle model (1) in the main text, it is as shown in **Supplementary Figure 17a**, which is different from **Fig. 3e** (**Supplementary Figure 17b**). To explain the difference, let us consider the large D>0 case on the hole side (i.e., the upper left part of **Supplementary Figure 17 a,b**).

The single particle picture of ZLL in BBG with small Zeeman energy expects the filling sequence from $\nu = -4$ should be $|N\xi\sigma\rangle=|0\text{-}\downarrow\rangle,|0\text{-}\uparrow\rangle,|1\text{-}\downarrow\rangle,|1\text{-}\uparrow\rangle$ for large D>0 case. However, Coulomb interactions change this filling order. Coulomb interaction favors sequential filling of N=0,1 with same spin due to the exchange interaction.[7] i.e. filling different orbitals with same spin,(e.g. $|0\text{-}\downarrow\rangle,|1\text{-}\downarrow\rangle$) has more favorable Coulomb energy than filling same orbital of opposite spin(e.g. $|0\text{-}\downarrow\rangle,|0\text{-}\uparrow\rangle$).

The experimental filling sequence(orbital, valley, spin) of ZLL in our sample was identified by analyzing the Landau fan diagram and n–D sweep. First, we analyzed the orbital nature(N=0 or 1) of the hole side of ZLL at the large D>0 case(i.e. the upper left side of **Fig. 3c**). By observing the ZLL in **Fig. 2a** or **5a**, we can see that filling -4<$\nu$<-3 looks similar to -2<$\nu$<-1(both shows conventional fractional quantum Hall state up to $\phi/\phi_0 = 1/3$), while -3<$\nu$<-2 looks similar to -1<$\nu$<0 (a similar moiré Chern band structure emerged from $\phi/\phi_0 \approx 0.2$). This similarity is due to the same orbital nature of each pair[8] and the single particle calculation in ref (6)[6] showed that the moiré Chern band was developed at a lower magnetic field for N=1 state. This supports the filling order $|N\xi\sigma\rangle=|0\text{-}\downarrow\rangle,|1\text{-}\downarrow\rangle,|0\text{-}\uparrow\rangle,|0\text{-}\uparrow\rangle$ for large D>0 case in our sample(large D makes valley $\xi = -$ fills before valley $\xi = +$). And in the reference cited in the main text[9], a detailed experimental and theoretical account of the ZLL in the absence of a moiré potential was provided. In ref.(5), the filling sequence of ZLLs according to the size of the interaction is simulated, and it can be seen that in the weak interaction regime, the ZLLs are filled in the order of N=0,0,1,1 as discussed above, and in the intermediate and strong interaction regime, the ZLLs are filled in the order of N=0,1,0,1. By tracking two states feeling moiré potential strongly($|N\xi\sigma\rangle=|1\text{-}\downarrow\rangle,|1\text{-}\uparrow\rangle$) in **Fig. 3c**, we can see that our data is more consistent with the intermediate interaction regime confirmed experimentally in ref.(9).

Additionally, if the magnetic field is further increased, the filling sequence mentioned above can be flipped once more. Since $\Delta_{10}$is proportional to B, while the Coulomb interaction $E_C$ is proportional to $\sqrt{B}$, there will be a point where $\Delta_{10} > E_C$ as the magnetic field increases. In this case, the single particle effect will dominate again, and the electrons will be filled in order $|0\text{-}\downarrow\rangle,|0\text{-}\uparrow\rangle,|1\text{-}\downarrow\rangle,|1\text{-}\uparrow\rangle$. However, in our experimental setup, we did not reach the magnetic field where this reversal occurs

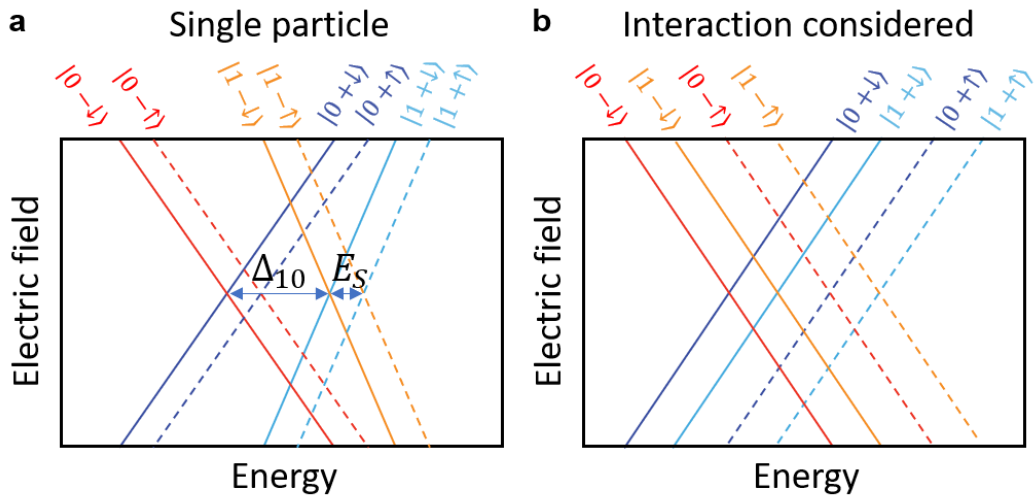

**Supplementary Figure 17 |** Schematic of ZLL of BBG energy spectrum in purely single particle picture (**a**) and interaction considered (**b**). Note that **b** is purely illustrative since the actual energetics, taking into account interactions, do not simply come out as a straight line. Also, the specific energy gap and slope differences were ignored in **b**.

**Supplementary Note 8. Magnetotransport data from other devices.**

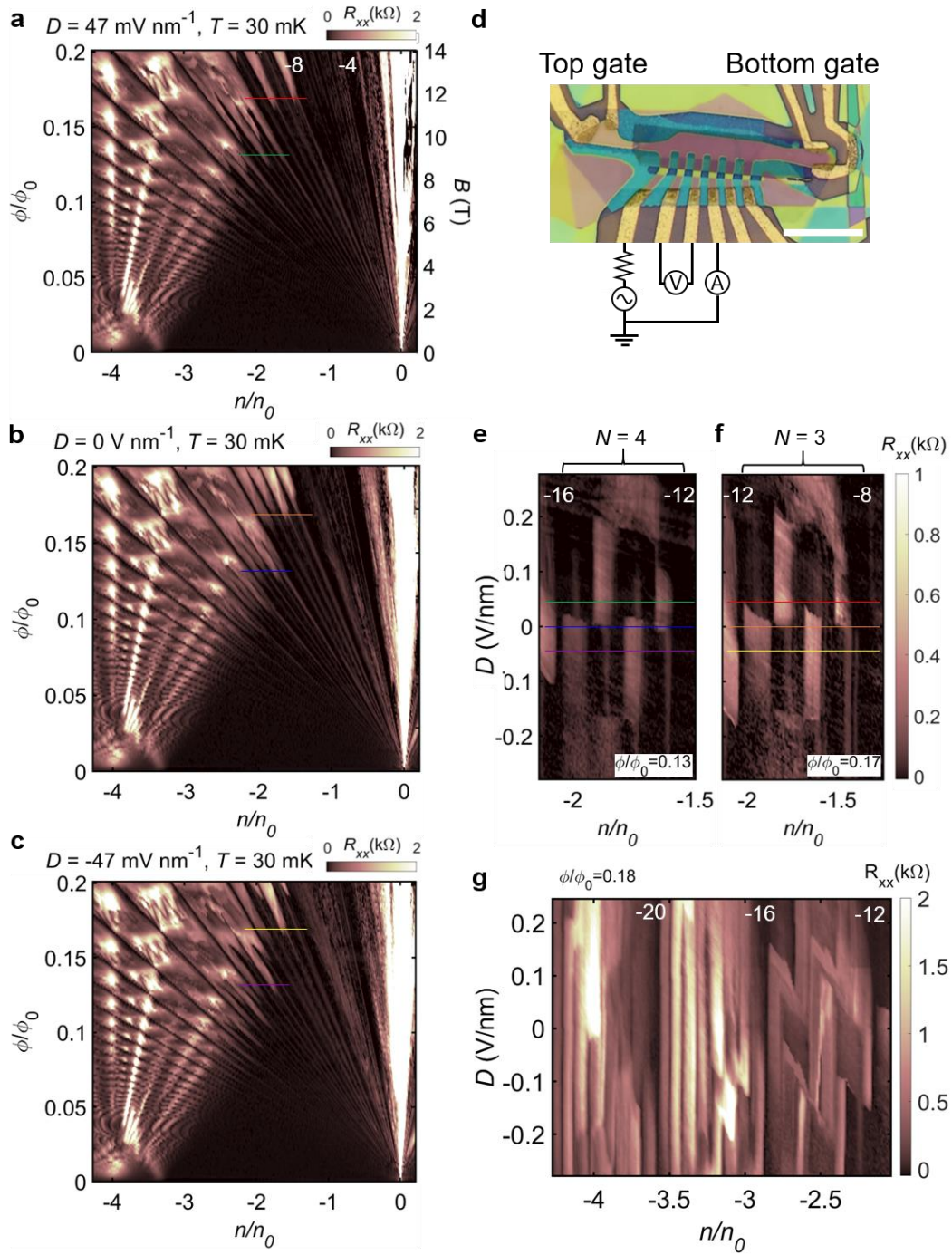

**Supplementary Figure 18 | Magnetotransport measurement of 1.37 deg sample. a–c,** Landau fan diagram of longitudinal resistance up to B=14T at different vertical displacement fields. All measurements were conducted at T=30mK. Scale bar, 15 μm. **d,** Optical microscope image of the measured device and the schematic of measurement configuration. **e–g,**

Longitudinal resistance $R_{xx}$ as a function of the carrier density n and D field at 8.9T(**e**), 11.75T (**f**), 12.4T (**g**). N denotes the orbital number of the Landau level of BBG and white numbers in **a**, **d** denote the Landau filling factor. Each colored horizontal line in **e–f** corresponds to the same colored lines in **a–c**.

We fabricated a 1.37 deg aligned device similar to the device used in the main text. (top hBN: 67 nm, bottom hBN: 50 nm) The sample showed similar valley-selective moiré effects in each Landau level and D field tunability of Chern insulators. We observed integer quantum Hall, Chern insulator, and fractional quantum Hall (up to N=3 LL) states in the Hofstadter butterfly pattern of the new sample, but not symmetry broken Chern insulator and fractional Chern insulator states. For the device of the main text, interacting states appeared in the low (less than $1 \times 10^{12} \ cm^{-2}$) carrier density region, but the new device required about three times higher carrier densities to fill the same superlattice filling factor, so we could observe only the high density and low $\phi/\phi_0$ region of Hofstadter patterns in the magnetic field range of our system. We believe that this device will also show interaction driven states if the magnetic field is further increased to access the low carrier density region of the Hofstadter pattern.

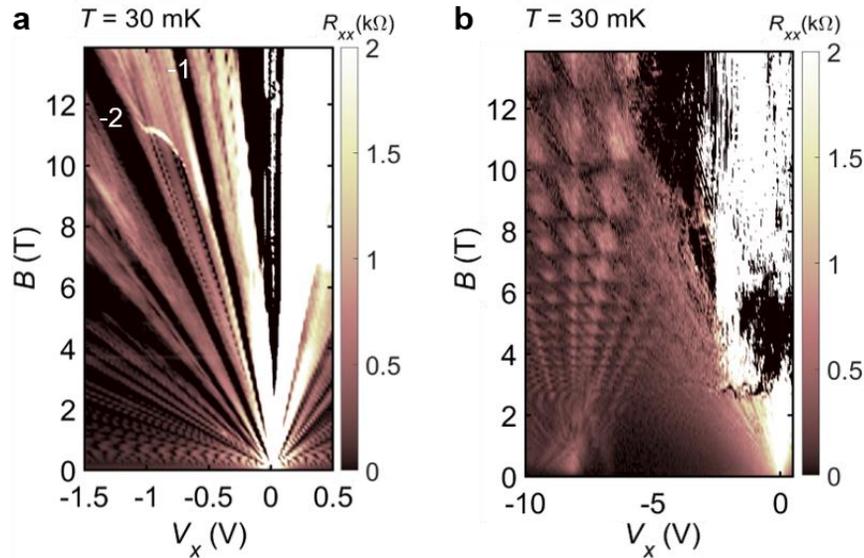

**Supplementary Figure 19 | Magnetotransport measurement of other samples. a**, Device with gate leakage. Only a small range of carrier density was tuned, and a transition to higher resistance in the ZLL metallic state was observed around B=11T. White number denotes the Landau filling factor of BBG. **b**, Device with bad contact resistance. Fabricated without contact graphite and using a flake with naturally connected trilayer graphene and BBG regions, the contact quality became bad with increasing magnetic field.

**Supplementary Note 9. Helical edge states at t=0 insulators.**

The left column of **Supplementary Figure 8** below shows the longitudinal resistance ($R_L$) measurement, and the right column shows the non-local resistance ($R_{NL}$) measurement. When comparing the two data sets, the state (0,-1) is particularly standing out with unusually high values of $R_{NL}$ over the corresponding $R_L$.

Our interpretation of the data is based on the following analysis: (We believe this is also consistent with the observation of Sanchez-Yamagishi et al. Nature nanotechnology 12 (2), 118-122 (2017))[10]

1. Compressible state: Intermediate $R_L$, very low $R_{NL}$
   - Reasoning: The voltage drops for $R_{NL}$ will be decaying far from the source-drain electrodes.
2. Incompressible state without an edge channel: High $R_L$, very low $R_{NL}$
   - Reasoning: The voltage drops for $R_{NL}$ will be decaying far from the source-drain electrodes.
3. Incompressible state with chiral edges: Zero $R_L$, zero $R_{NL}$
   - Reasoning: Chiral edges would act as equipotential 1D channels to give nearly zero NL resistance
4. Disordered inhomogeneous phase of multiple states: Intermediate $R_L$, low $R_{NL}$
   - Reasoning: Inevitable dissipations occur at every boundary of different states and thus make the situation qualitatively similar to the one of the compressible states.
5. Incompressible state with helical edges: a sizable fraction of the quantum conductance for both $R_L$ and $R_{NL}$
   - Reasoning: The voltage drops occur near ohmic contacts due to selective mixing of helical 1D channels.

As expected from the above analysis, in **Supplementary Figure 8d**, most compressible states and chiral incompressible states display highly suppressed $R_{NL}$. Looking at the states of $t = 0$, we can see that they generally have sizable $R_{NL}$'s. The $R_{NL}$ for (t,s)=(0,-4) is around 1 k$\Omega$ at all magnetic field values. The $R_{NL}$ for (0,-2) is similar in size, except for some regions near $\phi/\phi_0 \sim 0.2, 0.4$. For (0,-1), while the longitudinal resistance peak is the smallest of the three, $R_{NL}$ shows the largest value, approaching $h/4e^2$ at high magnetic fields, which is the expected value to be measured for states with a single helical edge when the NL configuration is used as shown in **Supplementary Figure 8b**, based on the Landauer–Büttiker formalism. This strongly suggests the existence of non-local helical transport at (0,-1). In the same vein, the state at (0,-2) near $\phi/\phi_0 \sim 0.4$ also hints on a helical state.

For the question about the nature of the state that can harbor a helical edge in this system, we think it is a state having the valley-dependent Chern numbers with opposite signs at K and K', with suppressed intervalley scattering. Actually, our calculation of the valley-specific Hofstadter spectra in **Supplementary Figure 16** shows that, in ZLL, where the observed (0,-1) state resides, the K valley has a Chern number of -2 over a wide range of energy (the absence of states for K) while the K' valley displays moiré-induced subbands with many different Chern numbers. From the fact that the total Chern number of the observed (0,-1) state should be 0 means that an energy gap opens at the Fermi level with the Chern number of +2 in the K' valley. We originally named this state as a QVH state, in which the valleys have edge channels with opposite chirality, forming helical edge states.

We note that, since the spectra in **Supplementary Figure 16** are limited to the spin degenerate case, a spin hall state or a complex spin–valley magnetic state are not considered and thus cannot be ruled out given the spin degeneracy is to be broken in the real system. Nonetheless, we think that the valley-dependent Chern number provides a crucial mechanism in the formation of the helical edge state.

**Supplementary Note 10. Temperature dependence of $n/n_0$=-4 satellite peak at zero magnetic field.**

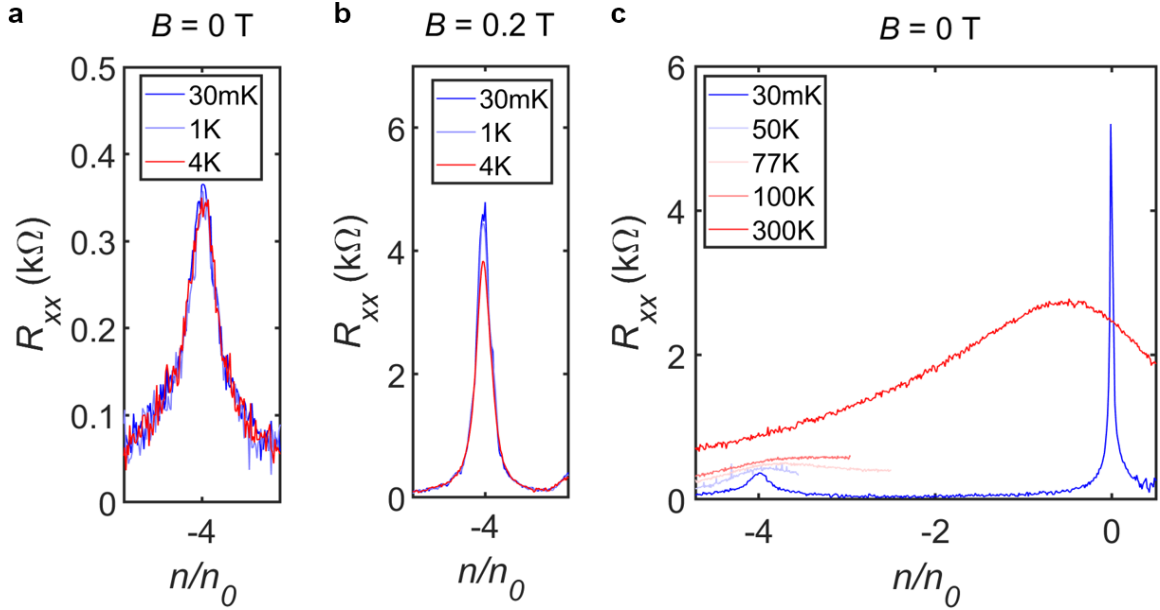

**Supplementary Figure 20 | Temperature dependent longitudinal resistance measurement at $D$=0. a**, $R_{xx}$ peak of $n/n_0 = -4$ at $B$=0T up to $T$=4K. **b**, $R_{xx}$ peak of $n/n_0 = -4$ at $B$=0.2T up to T=4K. In the presence of a magnetic field, the $n/n_0 = -4$ peak shows insulating behavior. **c**, High temperature $R_{xx}$ measurements at B=0 up to T=300K

Theoretical papers on BBG–hBN aligned devices predict either a full gap or a DOS minimum at the secondary Dirac point resulting from the moiré pattern, depending on the parameters used. A recent paper on a small angle twisted BBG–hBN system shows that the band structure with a DOS minimum is in better agreement with experimental results. To verify whether the satellite peak in our device is a gap or a DOS minimum, we performed transport measurements at higher temperatures (**Supplementary Figure 20c**). Up to 4 K, where most of the experiments in the main text were performed, it is difficult to distinguish whether the satellite peak is metallic or insulating in the absence of a magnetic field. (**Supplementary Figure 20a**) However, as we increase the temperature beyond a few tens of K, we find that the resistance of the satellite peak increases with it, confirming that it is a DOS minimum rather than a full gap at n/n_0=-4 in our sample.

**Supplementary Note 11. Characterization of graphite contacts.**

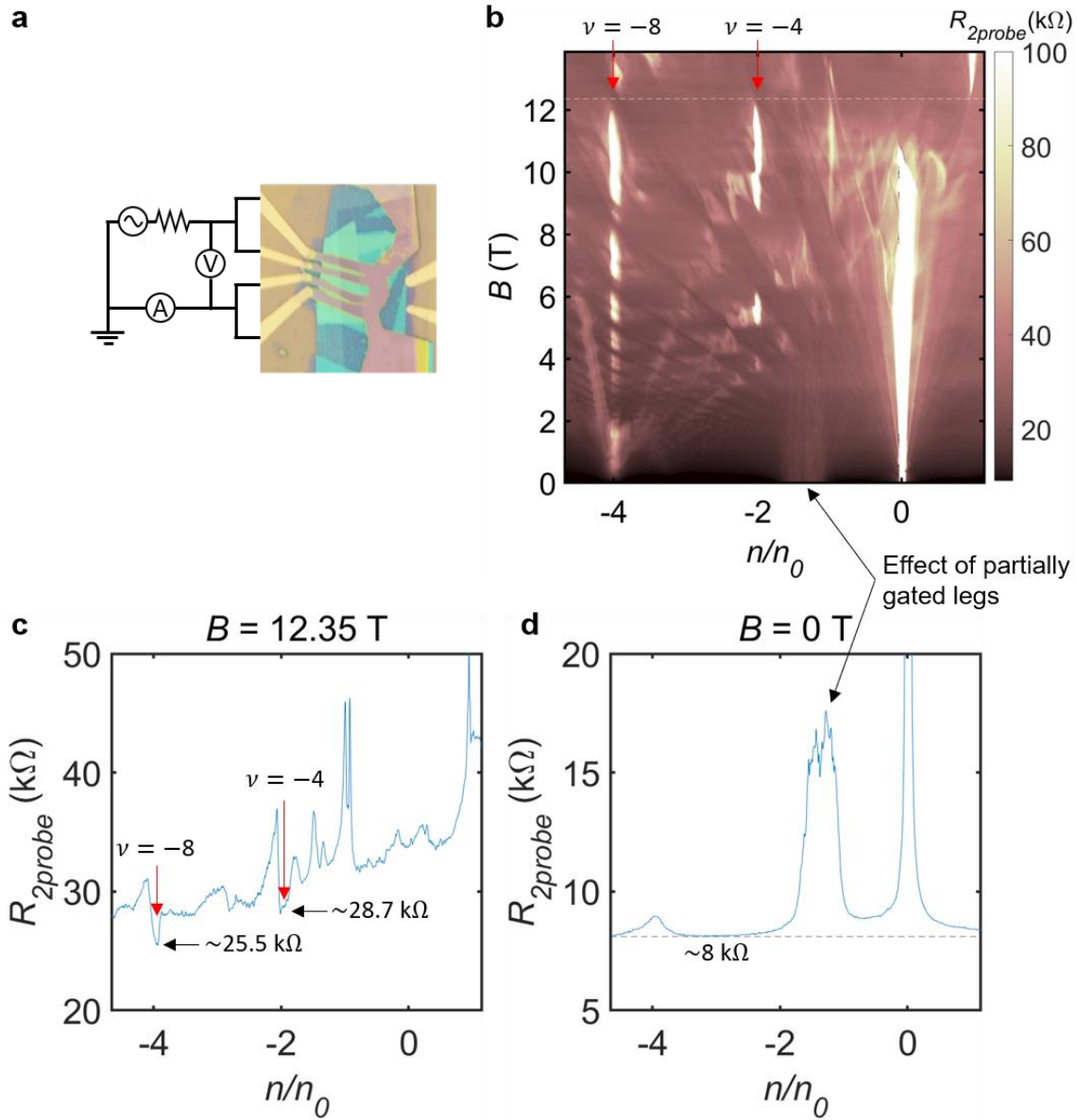

**Supplementary Figure 21 | Contact resistance characterization. a,** Measurement configuration of two-probe resistance. Two legs are paired in parallel and the series resistance of the two pairs was measured. This makes the measured quantity approximate the resistance coming from a single leg. **b,** Landau fan diagram of 2-probe resistance at $T = 30$ mK, $D$=116 mV/nm$^{-1}$. The colormap is truncated at both ends. **c,** Line cut at $B$=12.35T ($\phi/\phi_0 = 0.5$, white dashed line in **b**)。 The red arrows in **b,c** indicate Landau filling factor -4 and -8 points. **d,** Line cut at B=0T. The 8 kΩ line is shown as a black dashed line. The black arrows denote the area affected by the partially gated legs.

To characterize the contact resistance, we conducted an additional two-probe magnetoresistance measurement at $T = 30$ mK. At zero field, the lowest value of the two-probe resistance was measured to be about 8 kΩ, which directly gives an approximate contact resistance due to the small four-probe resistance. At high magnetic fields, we compared the resistance values of the ν=-4 and ν=-8 IQHEs, which are among the strongest features in the single particle picture, and found them to be about 28.7 kΩ and 25.5 kΩ, respectively. We analyzed that the difference, about 3.2 kΩ, came from the difference in conductance of the edge states (($h/e^2$)/8 ~ 3.2 kΩ), so we estimated the residual resistance to be about 22 kΩ. Subtracting 2 kΩ of cryostat wire resistance from this, the contact resistance is approximately 6 kΩ at zero field and 20 kΩ at 12.35 T. (This is not much different at 14 T) Although it increases by about a factor of 3, we found the graphite contact remained good enough that even the two-probe measurement was able to reproduce most of the detailed features observed in the four-probe measurement.

**Supplementary References**